\documentclass[usegraphicx,useAMS,usenatbib]{mn2e}
\usepackage{times}
\usepackage{amssymb}
\usepackage{amsmath}
\usepackage{graphicx}
\usepackage{euscript}
\usepackage{rotating}
\usepackage{epsfig}
\usepackage{url}

\def\ghz{{\rm\thinspace GHz}}

\def\kpc{{\rm\thinspace kpc}}
\def\kmpspmpc{\hbox{$\rm\thinspace km~s^{-1}~Mpc^{-1}$}}
\def\kev{{\rm\thinspace keV}}
\def\ergps{{\rm\thinspace erg~s^{-1}}}

\def\gyr{{\rm\thinspace Gyr}}
\def\yr{{\rm\thinspace yr}}
\def\kilos{{\rm\thinspace ks}}
\def\ergpcmsqps{{\rm\thinspace erg~cm^{-2}~s^{-1}}}

\begin{document}

\title{Investigating Heating and Cooling in the BCS \& B55 Cluster Samples}
\author[Dunn \& Fabian]
{\parbox[]{6.in} {R.J.H. Dunn$^1$\thanks{E-mail: r.j.dunn@phys.soton.ac.uk} and A.C.Fabian$^2$\\
    \footnotesize
    $^1$ School of Physics and Astronomy, University of Southampton,
    Southampton, SO17 1BJ\\$^2$ Institute of Astronomy, Madingley
    Road, Cambridge, CB3 0HA}}

\maketitle

\begin{abstract}
We study clusters in the BCS cluster sample which are observed by {\it
  Chandra} and are  more
distant than redshift, $z>0.1$.  We select from this subsample the clusters which have both a short central cooling time and a
central temperature drop, and also those with a central radio source.
Six of the clusters have clear bubbles near the centre.  We calculate
the heating by these bubbles and express it as the ratio $r_{\rm heat}/r_{\rm cool}=1.34\pm0.20$.  This result is used to
calculate the average size of bubbles expected in all clusters with central
radio sources.  In three cases the predicted bubble sizes approximately match the
observed radio lobe dimensions.  

We combine this cluster sample with
the B55 sample studied in earlier work to increase the total sample
size and redshift range.  This extended sample contains 71
clusters in the redshift range $0\leq z \leq0.4$.  The
average distance out to which the bubbles offset the X-ray cooling in
the combined sample is
at least $r_{\rm heat}/r_{\rm cool}=0.92\pm0.11$.  The
distribution of central cooling times for the combined sample shows no clusters with clear
bubbles and $t_{\rm cool}>1.2 \gyr$.  An investigation of the
evolution of cluster parameters within the redshift range of the
combined samples does not show any clear variation with redshift.

\end{abstract}


\section{Introduction}\label{sec:intro}

Since the discovery of ``holes'' in the X-ray emission at the centre
of the Perseus Cluster \citep{Bohringer93}, and especially since the
launch of \emph{Chandra} and \emph{XMM-NEWTON}, many depressions
in the central intra-cluster medium (ICM) of low redshift clusters have been
found (e.g. Hydra A, \citealp{McNamara00}; A2052,
\citealp{Blanton01}; A2199, \citealp{Johnstone02}; Centaurus,
\citealp{Sanders02}).  Recent compilations are given by
\citet{Dunn06f,Rafferty06} and \citet{Birzan04}.  These holes have been observed to
anti-correlate spectacularly with the radio emission from the active
galactic nucleus (AGN) at
the centres of these clusters.  Their morphology, particularly
in the closest clusters, has led to their interpretation as
bubbles of relativistic gas blown by the AGN into the thermal ICM.
The relativistic gas is less dense than the ICM, so the bubbles
detach from the core and rise up buoyantly through the
cluster, e.g. Perseus \citep{Churazov00,Fabian03b}.  The older bubbles tend not to have
$\ghz$ radio emission associated with them and have been termed ``Ghost'' bubbles.

The X-ray emission of the ICM represents an energy loss, so leads the
plasma cooling in the core if there were no compensating energy source.  To maintain  pressure
support, the gas would flow on to the
central galaxy as a ``cooling flow.''  However, with the high spatial and
spectral resolution of \emph{Chandra} and \emph{XMM-Newton}, the gas
temperature is seen to drop by a factor of several in the core but no
continuous cooling flow is found (\citealp{Peterson03}, see \citealp{Peterson06}
for a review).  Some form of gentle,
continuous and distributed heating appears to be at work
\citep{Voigt04}.  The interaction of the AGN with the ICM is the
likeliest explanation as the heat source for cool-core clusters.  The
bubble model for radio sources was proposed by \citet{Gull73}, and
further theoretical results were presented in \citet{Churazov00,
  Churazov02}.  Recent studies on the heating ability of AGN within
clusters are \citet{Birzan04, Rafferty06} and \citet{Dunn06f}.

A majority (71 per cent) of ``cooling core''  clusters harbour radio
sources \citep{Burns90}, and a similar number of clusters which
require heating (likely to be a cooling core) harbour clear bubbles
\citep{Dunn05, Dunn06f}.  The action of creating the bubbles at the centre of the cluster by the
AGN is a favoured method of injecting energy into the central regions
of the cluster and so prevent the ICM from cooling.  This process sets
up sound/pressure waves in the ICM, the dissipation of which requires the ICM is to be
viscous \citep{Fabian03a}.  The viscous dissipation of the
pressure waves allows the energy from the bubble
creation to be dissipated far from the cluster centre where it is required.

In \citet{Dunn06f} we analysed clusters in the Brightest 55 (B55)
sample \citep{Edge90}.  These are all at comparatively low redshift
(only two are at $z>0.1$).  Using the results of \citet{Peres98} and
the literature at the time we selected those clusters with both a
short central cooling time and a large central temperature drop, as
well as those with a central radio source.  We found that at least 70 per cent of clusters in which high
cooling rates are expected (those with short central cooling times and
central temperature drops) harbour AGN bubbles.  We also found that
95 per cent of these cool core clusters harbour radio cores.  The total fraction of
clusters which harbour a radio source is 53 per cent (29/55).  Where AGN
bubbles are present, on average they show that the AGN is injecting
enough energy into the central regions of the cluster to offset the
X-ray cooling out to the cooling radius ($r_{\rm cool}$ where $t_{\rm
  cool}=3\gyr$).  To increase the number of clusters studied and also
investigate any evolution in the AGN heating with redshift we now turn to
the Brightest Cluster Sample \citep{Ebeling98}, initially performing a similar analysis
and then combining the results from both cluster samples.

The sample selection for this work is outlined in Section
\ref{sec:selection} and the data preparation and reduction in Section
\ref{sec:data_prep}.  We then look at the two different cluster
subsets, those with bubbles in Section \ref{sec:bubbles} and those
with central radio sources in Section \ref{sec:radio}.  The results of
our investigations into the global differences within the different
subsets is presented in Section \ref{sec:heating}.  The combined
cluster sample and the redshift evolution results are presented in
Section \ref{sec:discuss}.  We use $H_0=70\kmpspmpc$ with $\Omega_{\rm
  m}=0.3$, $\Omega_\Lambda=0.7$ throughout this work.

\section{Sample Selection}\label{sec:selection}

\begin{table*}
\caption{\label{Tab:clusterlist} {\sc Cluster Sample}}
\begin{tabular}{llrrccccc}
\hline
\hline
Cluster&Redshift&ObsId&Exposure (ks)&$t_{\rm cool}$&$T$ drop&Both&Bubbles&Radio\\
\hline
         A68 &  0.255 &  3250 &    9.9 &       &       &       &       &       \\ 
      A115 N &  0.197 &  3233 &   49.7 &     Y &     Y &     Y &     Y &     Y \\ 
      A115 S &  0.197 &  3233 &   49.7 &       &     Y &       &       &       \\ 
        A267 &  0.227 &  1448 &    7.4 &       &       &       &       &       \\ 
        A520 &  0.202 &   528 &    9.5 &       &     Y &       &       &       \\ 
        A586 &  0.171 &   530 &   10.0 &       &       &       &       &     Y \\ 
        A665 &  0.182 &   531 &    8.9 &       &       &       &       &       \\ 
        A697 &  0.282 &  4217 &   19.5 &       &       &       &       &       \\ 
        A773 &  0.217 &  5006 &   19.8 &       &       &       &       &       \\ 
        A781 &  0.298 &   534 &    9.9 &       &     Y &       &       &     Y \\ 
        A963 &  0.206 &   903 &   35.9 &       &     Y &       &       &       \\ 
       A1068 &  0.139 &  1652 &   25.5 &     Y &     Y &     Y &       &     Y \\ 
       A1201 &  0.169 &  4216 &   24.0 &       &       &       &       &       \\ 
       A1204 &  0.171 &  2205 &   23.3 &     Y &       &       &       &       \\ 
       A1413 &  0.143 &  5003 &   75.0 &       &       &       &       &       \\ 
       A1423 &  0.213 &   538 &    9.7 &     Y &     Y &     Y &       &     Y \\ 
       A1682 &  0.226 &  3244 &    5.4 &       &     Y &       &       &     Y \\ 
     A1758 N &  0.279 &  2213 &   49.7 &       &       &       &       &     Y \\ 
       A1763 &  0.223 &  3591 &   19.5 &       &       &       &       &     Y \\ 
       A1835 &  0.252 &   495 &   19.3 &     Y &     Y &     Y &     Y &     Y \\ 
       A1914 &  0.171 &  3593 &   18.8 &       &       &       &       &       \\ 
       A2111 &  0.229 &   544 &   10.2 &       &     Y &       &       &       \\ 
       A2204 &  0.152 &   499 &   10.1 &     Y &     Y &     Y &       &     Y \\ 
       A2218 &  0.171 &  1666 &   40.1 &       &       &       &       &       \\ 
       A2219 &  0.228 &   896 &   42.3 &       &       &       &       &     Y \\ 
       A2259 &  0.164 &  3245 &   10.0 &       &       &       &       &       \\ 
       A2261 &  0.224 &  5007 &   24.3 &     Y &       &       &       &       \\ 
       A2294 &  0.178 &  3246 &    9.0 &       &     Y &       &       &       \\ 
       A2390 &  0.231 &  4193 &   89.8 &     Y &     Y &     Y &       &     Y \\ 
       A2409 &  0.147 &  3247 &   10.2 &       &       &       &       &       \\ 
       HercA &  0.155 &  6257 &   49.4 &     Y &     Y &     Y &     Y &     Y \\ 
 MS0906+1110 &  0.175 &   924 &   29.6 &       &       &       &       &       \\ 
 MS1455+2232 &  0.258 &  4192 &   91.4 &     Y &     Y &     Y &       &     Y \\ 
RXJ0439+0520 &  0.208 &  1649 &    9.6 &     Y &     Y &     Y &       &     Y \\ 
RXJ0439+0715 &  0.245 &  3583 &   19.2 &       &       &       &       &     Y \\ 
RXJ1532+3021 &  0.362 &  1649 &    9.2 &     Y &     Y &     Y &     Y &     Y \\ 
RXJ1720+2638 &  0.161 &  4361 &   23.3 &     Y &     Y &     Y &       &     Y \\ 
RXJ2129+0005 &  0.234 &   552 &    9.9 &     Y &     Y &     Y &       &     Y \\ 
    ZwCl1953 &  0.373 &  1659 &   20.7 &       &       &       &       &     Y \\ 
    ZwCl2701 &  0.214 &  3195 &   26.7 &     Y &     Y &     Y &     Y &     Y \\ 
    ZwCl3146 &  0.285 &   909 &   45.6 &     Y &     Y &     Y &     Y &     Y \\ 
    ZwCl5247 &  0.229 &   539 &    9.0 &       &       &       &       &     Y \\ 
\hline
Totals & & & 42& 16& 21& 14& 6& 23\\
\hline
\end{tabular}
\begin{quote} All the clusters in the sample, for the sample
  selection see text.  The exposure time is that after reprocessing
  the data.  The cooling times and central temperature drops have been
  determined from the profiles calculated during the course of this
  work.  Clusters with $<3\gyr$ are classed as having a short central
  cooling time, and those
  with $T_{\rm centre}/T_{\rm outer}<1/2$, a central temperature drop.
  The radio detections come from the NVSS.
\end{quote}
\end{table*}

To increase the number of clusters studied and to investigate the redshift evolution of AGN in clusters we now
turn to the {\it ROSAT} Brightest Cluster Sample (BCS, \citealp{Ebeling98}).  Most of the
B55 clusters are at low redshifts ($z\lesssim 0.1$), whereas the BCS
clusters extend to higher redshifts.  The BCS is a 90 per cent
flux-complete sample comprising the 201 X-ray-brightest clusters in
the northern hemisphere at high galactic latitudes ($|\,b\,|>20^\circ$ \citealp{Ebeling98}).
The member
clusters have measured redshifts of $z\lesssim0.4$ and fluxes greater than
$4.4\times 10^{-12}\ergpcmsqps$ in the $0.1-2.4\kev$ band (for
$H_0=50\kmpspmpc$).  There is a low flux extension to the BCS, the eBCS, which includes clusters with
fluxes in the range $2.8\times 10^{-12}$ to $4.4\times 10^{-12}\ergpcmsqps$
\citep{Ebeling00}.  \citet{Bauer05} investigated the existence of cool cores
in the high redshift end of the BCS ($z>0.15$), and found that 34 per
cent of clusters exhibited signs of strong cooling (central $t_{\rm
  cool}<2\gyr$).  In our study ($z>0.1$ and $t_{\rm
  cool}<3\gyr$) we find that 16 clusters have a short central cooling
time, corresponding to 38 per cent.  If the temperature drop is also
required for ``strong cooling'' then the fraction drops to 33 per
cent, both of which match their result.  Globally there have been a
variety of estimates on the fraction of clusters which have cool
cores\footnote{The ones which would previously have been described as
  having a ``cooling flow''.}  \citet{Peres98} used the B55 sample and
found a fraction of 70-90 per cent.  However
they define a ``cooling flow'' cluster is that which has
$t_{\rm cool}<t_{\rm cluster}=13\gyr$.  If we use $t_{\rm
  cool}<13\gyr$ for our selection of cool core clusters, 38/42
(90 per cent) of clusters have a cool core.  It is therefore
reasonable to assume that the {\it Chandra} observations of clusters
within the BCS have not been highly biased towards cool core
clusters.  In the B55 sample from
\citet{Dunn06f} only 1/30 clusters would not have a cool core, though the
selection procedure was different in that study.

So that the BCS and B55 cluster samples do not overlap in this study, we only select
those clusters in the BCS with $z>0.1$ as the parent sample for this
study.  There is only one cluster in common with both the B55 and BCS
with this cutoff, A2204.  There are 87 clusters
(including A2204) in the parent sample, of which 42 have 
observations in the {\it Chandra} archive.  These are shown in Table
\ref{Tab:clusterlist}.  We assume that the {\it Chandra} observations
have not been biased to a particular type of cluster\footnote{Only
  around 1/3 of the clusters observed with {\it Chandra} would be
  classed as cool-core clusters.  A quick analysis of the average
  length of observations for different cluster types is discussed in
  Section \ref{sec:heating}} and so quote
population fractions from this sample of 42 rather than from the parent
sample of 87 clusters.

Following the data reduction (see Section \ref{sec:data_prep}) this parent
sample was further split into sub-samples, following \citet{Dunn06f}:
those clusters which harbour clear bubbles; those in which
heating is required to prevent rapid cooling and
those which harbour a radio source were identified.  The allocation of
clusters into the three sub-samples is shown in Table \ref{Tab:clusterlist}.

To identify those clusters which are likely to require some form of
heating to prevent large quantities of gas from dropping out we take
those which have a short central cooling time ($<3\gyr$) and a
large central temperature drop ($T_{\rm centre}/T_{\rm outer}<1/2$).
Exactly half (21) of the clusters in the sample have a central
temperature drop, and 16 have a short central cooling time (16/42, 38 per cent, similar to
\citealt{Bauer05}, although they use $t_{\rm cool}<2\gyr$).  14 clusters have both (14/42, 33 per
cent), of which at least 6 have clear bubbles (6/14, 43 per cent;
6/42, 14 per cent).  We use the NVSS\footnote{NRAO (National Radio
  Astronomy Observatory) VLA (Very Large Array) Sky Survey} to
determine whether the clusters have a central radio source.  23
clusters harbour a central radio source (23/42, 55
per cent), including all those which require some form of heating.
There appears to be a category of clusters which have cool cores and a
radio source but in which no clear bubbles have been detected.  Both
in the B55 sample (Table 2 \citet{Dunn06f}, $2^{\rm nd}$ column) and
the BCS have these sorts of clusters, with similar fractions in each
(5/30 in the B55 and 8/42 in the BCS).

Comparing with the B55 sample, \citet{Dunn06f} found that 
the fraction of clusters which require
heating is 25 per cent, and which harbour a
radio source is at least 53 per cent.  Of
the clusters which require some form of heating, at
least 70 percent harbour clear bubbles and 95 percent harbour a
central radio source.  

In comparing the number of identified bubbles in the BCS and B55
sample it is expected that as a result of the greater distance of
clusters in the BCS, fewer bubbles would be identified even if the
same proportion of clusters harboured them.  It is therefore not
surprising that the percentage of clusters which harbour bubbles is
half that determined for the B55 (from those clusters which require
heating).  The proportion of clusters which harbour radio sources, on
the other hand, is very similar (BCS=55, B55=53 per cent) as is the
proportion of clusters requiring heating (BCS=33, B55=25 per cent).

\begin{figure}
\centering
\includegraphics[width=0.95\columnwidth]{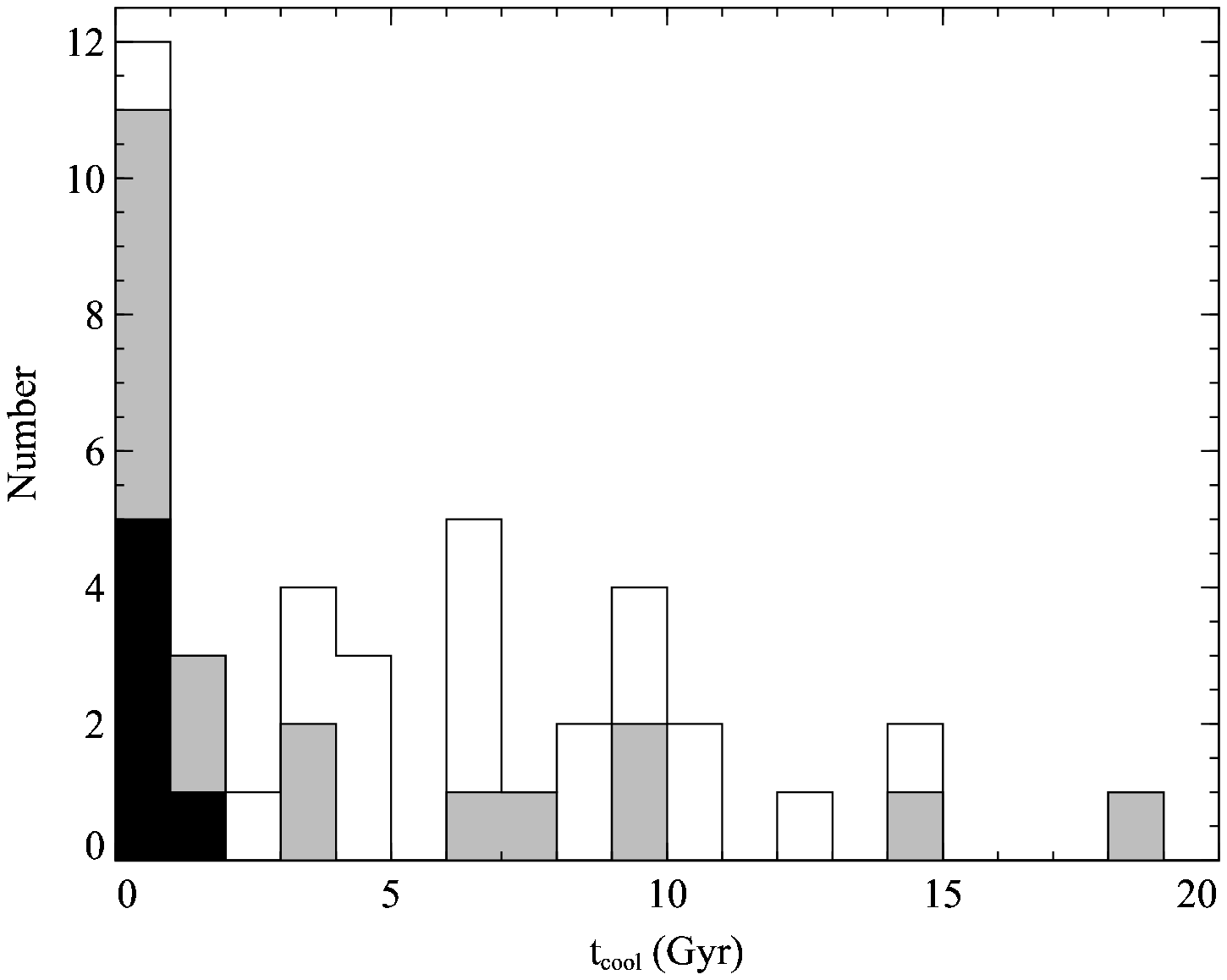}
\caption{\label{fig:tcool}The distribution of $t_{\rm cool}$ in the BCS
sample.  The black bars indicate those clusters which harbour clear
bubbles, and the grey ones, those with a central radio source.
ZwCl5247, which harbours a radio source, is beyond the right edge of
this plot.}
\end{figure}

\section{Data Preparation}\label{sec:data_prep}

The X-ray {\it Chandra} data of the clusters were
processed and cleaned using the CIAO software and calibration files
(CIAO v3.3, CALDB v3.2).
We began the reprocessing by removing the afterglow detection and
re-identifying the hot pixels and cosmic ray afterglows, followed by
the tool {\scshape acis\_process\_events} to remove the pixel randomisation and to
flag potential background events for data observed in Very Faint (VF)
mode.  The Charge-Transfer Efficiency was corrected for, followed by
standard grade selection.  Point-sources were identified using the
{\scshape wavdetect} wavelet-transform procedure.  For clusters observed with
the ACIS-S3 chip, the ACIS-S1 chip was used to form the light curves
where possible.  In all other cases, light-curves were
taken from on-chip regions as free as possible from cluster emission.
For the spectral analysis, backgrounds were taken from the CALDB
blank-field data-sets.  They
had the same reprocessing applied, and were reprojected to the correct
orientation.  

Cluster centroids were chosen to lie on the X-ray
surface brightness peak.  
Annular regions, centred on the centroids, were automatically assigned with constant
signal-to-noise, stopping where the background-subtracted surface brightness of the cluster dropped
below zero.   The initial signal-to-noise was 100, and this was
increased or decreased by successive factors of $\sqrt{2}$ to obtain a
number of regions between four and ten.  The minimum signal-to-noise
allowed was 10.  For clusters imaged on the ACIS-I chip, the effects
of excluding the chip-gaps was accounted for by adjusting the stop
angles of the annuli.

The $0.5-7\kev$
spectra were extracted, binned with a minimum of 20 counts per bin,
and, using {\scshape xspec} (v12.3)
(e.g. \citealt{Arnaud96}), a {\scshape projct} single temperature
{\scshape mekal} (e.g. \citealt{Mewe95}) model with a {\scshape phabs}
absorption was used to
deproject the cluster.  Chi-square was used as an estimator of the
spectral fit.  In some clusters the temperatures
for some of the regions were undefined.  To solve this problem, the minimum number of regions
was reduced, a maximum radius for the outermost annulus was set, or the outermost region was
removed to try to improve the behaviour of the profile.  Using the deprojected cluster temperature, abundance and normalisation profiles; density, pressure,
entropy, cooling time and heating profiles were calculated.  

These profiles give azimuthally averaged values for the cluster
properties and have been used in the subsequent calculations.  In some
clusters, for example M87 and Perseus, the central parts of the cluster
are not very smooth, e.g. due to bubbles.  \citet{Donahue06} show that
these features do not strongly bias estimates of the entropy.
As such the use of these azimuthally averaged values is not likely to
introduce large biases into the subsequent calculations.

\section{Clusters with Bubbles}\label{sec:bubbles}

The analysis of the clusters with clear bubbles is the same as in
\citet{Dunn06f}.  Where clear cavities are seen in the X-ray emission
then the energy injected into the central regions of the cluster is
calculated from the energy in the bubbles. 
\begin{equation}
E_{\rm bubble}=\frac{1}{\gamma_1-1}p_{\rm th}V+p_{\rm
  th}V=\frac{\gamma_1}{\gamma_1-1}p_{\rm th}V\label{Eq:E_bubble}
\end{equation}
where $V$ is the volume of the bubble, $p_{\rm th}$ is the thermal
pressure of the ICM at the radius of the bubble from the cluster
centre and $\gamma_1$ is the mean adiabatic index of the fluid in the
bubble.  Assuming the fluid is relativistic, then $\gamma_1=4/3$, and so
$E_{\rm bubble}=4p_{\rm th}V$.

We parametrise the bubbles as ellipsoids, with a semi-major axis
along the jet direction, $R_{\rm l}$, and a semi-major axis across it,
$R_{\rm w}$.  The depth of the bubbles along the line of sight is
unknown and so we assume them to be prolate ellipsoids, with the
resulting volume being $V=4\pi R_{\rm w}^2 R_{\rm l}/3$.  

To estimate the rate at which the AGN is injecting energy into the
cluster the we need to estimate the age of the bubble.
There are two principle methods of estimating the age of the bubble,
the sound speed timescale and the buoyancy rise time.  Observations of nearby
clusters have shown that young bubbles which are still being inflated do not create strong shocks
during their inflation.  They are therefore assumed to be expanding at
around the sound speed, and their age is $t_{\rm age}=t{c_{\rm
    s}}=R_{\rm l}/c_{\rm s}$ where the sound speed in the ICM is
\begin{equation}
c_{\rm s}=\sqrt{\gamma_2 k_{\rm B} T/\mu m_{\rm H}}\label{Eq:soundspeed}
\end{equation}
where $\gamma_2=5/3$.  As bubbles evolve they are expected to detach
from the AGN and rise up buoyantly through the ICM
\citep{Churazov00}.  Their age is best calculated by the buoyancy rise
time; $t_{\rm age}=t_{\rm buoy}=R_{\rm dist}/v_{\rm buoy}$, with the
buoyancy velocity given by
\begin{equation}
v_{\rm buoy}=\sqrt{2gV/SC_{\rm D}}\label{Eq:buoy}
\end{equation}
where $C_{\rm D}=0.75$ is the drag coefficient \citep{Churazov01} and
$S=\pi R_{\rm w}^2$ is the cross-sectional area of the bubble.

The radio sources in this cluster sample are analysed after the
identification of cavities within the X-ray emission.  We find two
strong radio sources within this sample, A115 (3C~28) and Hercules~A
(3C~348).  Both have current radio emission in a bi-lobed morphology,
and the radio lobes are still close to the core.  None of the others
have radio emission correlated with the cavities in the X-ray
emission.  We therefore assume that the radio emission has aged and so
faded beyond the sensitivity limit, and class these bubbles as Ghost bubbles.

Using the deprojected temperature, density and abundance profiles,
X-ray cooling was calculated for each spherical shell.  This was
compared to the power injected into the cluster by the AGN in the form
of bubbles ($P_{\rm bubble}=E_{\rm bubble}/t_{\rm age}$).  The
distance out to which the bubble power can counteract the X-ray
cooling was therefore calculated.  Uncertainties have been estimated using a simple Monte Carlo
simulation of the calculation\footnote{For this and the calculation of
  the expected bubble sizes in Section \ref{sec:radio}, the method is as follows.  The input
  values were assumed to have a Gaussian distribution.  The values
  used in each run of the simulation were then randomly selected.  After a large number of runs ($2\times10^5$) the
  resulting parameters were sorted into ascending order and the median
and interquartile range are used as the value and uncertainties.}.

\begin{table}
\caption{\label{Tab:bubbleprop} {\sc Clusters with Bubbles: Bubble Properties}}
\begin{tabular}{lllccc}
\hline
\hline
Cluster&Bubble&Type&$R_{\rm l}$&$R_{\rm w}$&$R_{\rm dist}$\\
&&&(\kpc)&(\kpc)&(\kpc)\\
\hline
     A115 &NE&Y&$32.1\pm3.2$&$28.78\pm2.8$&$39.0\pm3.9$\\
 &SW&&$27.2\pm2.7$&$20.3\pm2.0$&$45.2\pm4.5$\\
     A1835 &NE&G&$15.5\pm1.6$&$11.6\pm1.2$&$23.3\pm2.3$\\
 &SW&&$13.6\pm1.4$&$9.6\pm1.0$&$16.6\pm1.7$\\
     HercA &E&Y&$110\pm11$&$75.4\pm7.5$&$110\pm11$\\
 &W&&$124\pm12$&$75.4\pm7.5$&$124\pm12$\\
RXJ1532+30 &W&G&$23.5\pm2.4$&$23.5\pm 2.4$&$32.0\pm3.2$\\
ZwCl2701   &E&G&$29.8\pm3.0$&$29.8\pm3.0$&$42.6\pm4.3$\\
 &W&&$28.7\pm2.9$&$28.7\pm2.9$&$39.5\pm4.0$\\
ZwCl3146   &NE&G&$28.0\pm2.8$&$28.0\pm2.8$&$57.1\pm5.7$\\
 &SW&&$30.0\pm3.0$&$30.0\pm3.0$&$53.2\pm5.3$\\
\hline
\end{tabular}
\begin{quote}The bubbles are parametrised as ellipsoids, with $R_{\rm
    l}$ the semi-major axis along the jet direction,  and $R_{\rm w}$
  the semi-major axis across it.  We class young bubbles as those with
  clear radio emission, and ghost bubbles as those without.  Both A115
  (3C~28)
  and Hercules~A (3C~348) have strong bi-lobed radio sources at their centres.
\end{quote}
 \end{table}

As this cluster sample is at a higher average redshift than the
Brightest 55 sample studied in \citet{Dunn06f}, fewer clear bubbles
have been identified.  Also, the X-ray emission is fainter, and so
analysis of the outermost regions of the cluster becomes more
difficult.  Only large or extreme bubbles will be detected.  It is therefore not surprising that of the six clusters
which have clear depressions in the X-ray emission, in three of them
the AGN is injecting much more energy than is
being lost through X-ray emission in the central regions.  \citet{Nulsen05}
have studied the {\it Chandra} data of Hercules A and concluded that this
outburst is exceptionally powerful. A comparison between
the X-ray cooling luminosity within the cooling radius to the bubble
power is shown in Table \ref{Tab:bubblelist}.

\begin{table*}
\caption{\label{Tab:bubblelist} {\sc Clusters with Bubbles: AGN
    Heating ability}}
\begin{tabular}{lr@{$\;\pm\;$}lr@{$\;\pm\;$}lr@{$\;\pm\;$}llr@{$\;\pm\;$}lr@{$\;\pm\;$}l}
\hline
\hline
Cluster&\multicolumn{2}{c}{$L_{\rm  cool}^{\hspace{0.35cm} a}$}&\multicolumn{2}{c}{$4p_{\rm th}V/t_{\rm age}$}&\multicolumn{2}{c}{$r_{\rm cool}$}&$r_{\rm heat}$&\multicolumn{2}{c}{$r_{\rm heat}/r_{\rm
  cool}^{\hspace{0.35cm} b}$}&\multicolumn{2}{c}{Heated
  Fraction$^{c}$}\\
&\multicolumn{2}{c}{$(10^{43}\ergps)$}&\multicolumn{2}{c}{$(10^{43}\ergps)$}&\multicolumn{2}{c}{(\kpc)}&(\kpc)\\
\hline
{\bf A115 N} & $   16.8$&$  1.1$& $  273$&$   33$ & $   70.5$&$    3.8$ & $ >153$ & $   2.17$&$   0.12$ & $  16.4$&$   2.2$ \\ 
    A1835 & $  205$&$    6$& $  190$&$   38$ & $   96.7$&$    2.3$ & $  90.5\pm 14.9$ & $   0.94$&$   0.16$ & $   0.92$&$   0.19$ \\ 
{\bf HercA} & $    4.2$&$    0.4$& $ 1560$&$  360$ & $   45.9$&$    3.4$ & $> 600$& $  13.1$&$   1.0$ & $ 373$&$  98$ \\ 
RXJ1532+30 & $  218$&$   23$& $  331$&$  126$ & $  107$&$   10$ & $ 187\pm 96$&$ 1.75$&$   0.91$ & $   1.52$&$   0.60$ \\ 
{\bf ZwCl2701} & $   23.4$&$    1.9$& $  602$&$   82$ & $   67.4$&$    3.0$ & $> 301$ & $   4.47$&$   0.20$ & $  25.5$&$   4.1$ \\ 
{\bf ZwCl3146} & $  153$&$    7$& $  786$&$  144$ & $   93.0$&$    3.7$ & $> 500$ & $   5.38$&$   0.21$ & $   5.11$&$   0.97$ \\ 
\hline
\end{tabular}
\begin{quote} 
The clusters in bold only have a lower limit for $r_{\rm heat}/r_{\rm
  cool}$ as the AGN counteracts all the cooling from the X-ray
emission which could be analysed.
$^{a}$ $L_{\rm cool}$ is calculated for the $0.5-7.0\kev$ range.
$^{b}$ The radius $r_{\rm heat}$, as a fraction of the cooling radius $r_{\rm cool}$, out to
which the energy injection rate of the bubbles ($4p_{\rm th}V/t_{\rm age}$) can offset the X-ray
cooling within that radius.
$^{c}$ The fraction of the X-ray cooling that occurs within $r_{\rm
   heat}$.
\end{quote}
\end{table*}

The average distance, as a fraction of the cooling radius, out to
which the bubbles can offset the X-ray cooling calculated from the
results is presented in Table \ref{Tab:bubblelist}.  The uncertainty in
the mean was estimated using a simple bootstrapping method.  On
average, the clear bubbles in this sample counteract the X-ray cooling
out to $r_{\rm heat}/r_{\rm cool}=   1.34\pm   0.40$.  The fraction of the
X-ray cooling within the cooling radius offset by the action of the
AGN is $  1.22\pm   0.15$.  We emphasise that this average is only for
{\it two} clusters (A1835 and RXJ1532+30), and should be regarded as a
lower limit for the following reasons.

The values exclude
those clusters where the calculation only provides a lower limit on
the distance.  Adding these clusters into the calculation results in
$r_{\rm heat}/r_{\rm cool}= 4.31\pm1.10$ and a heated fraction of
$75.0\pm50.4$.  It is likely that, because of the greater average
distance of this cluster sample, that the clusters with clear bubbles
are those which are undergoing/have recently undergone \emph{powerful} AGN
outbursts.  The more gentle energy injection events would not be
easily identified in the X-ray emission.

We do not comment further on these results here, as there are only
six clusters in this sub-sample.  Further analysis, where these
clusters are combined with those from the B55 sample, is discussed in
Section \ref{sec:discuss}.

\section{Clusters with Radio Sources}\label{sec:radio}

All the clusters which require some form of heating harbour a radio
source, which is similar to the fraction found by \citet{Dunn06f} of
95 per cent.  Following the study in \citet{Dunn06f} we invert the
problem, and use the results from the clusters with bubbles on this
sample.  Using the average distance out to which the bubbles
counteract the X-ray cooling, we calculate the X-ray cooling luminosity for these clusters
and equate this to the energy injected by the AGN.
\[
P_{\rm bubble}= L_{\rm cool}(r<r_{\rm heat}),
\]
where $r_{\rm heat}$ is calculated from $r_{\rm heat}=
1.34r_{\rm cool}$.  We assume that these ``expected''
bubbles are young, and so are being inflated at the sound speed.
\[
R_{\rm bubble}=\sqrt{\frac{P_{\rm bubble}}{4 p_{\rm th} c_{\rm s}}}
\]
Uncertainties have been estimated using a simple Monte Carlo
simulation of the calculation.

\begin{table*}
\centering
\caption{\label{tab:hidden}{\sc Predicted Bubble Sizes}}
\begin{tabular}{lccccrc}
\hline
\hline
Cluster&$z$&$r_{\rm cool}$&\multicolumn{2}{c}{Predicted Bubble
  Radius}&Central&Radio dimensions\\
&&$(\kpc)$&$(\kpc)$&(arcsec)&X-ray S/N&(arcsec)\\
\hline
      A586 & 0.171& $  20.59\pm  10.63$ & $   2.39\pm   1.00$ & $   0.82\pm   0.34$ &     1.7  & -  \\ 
      A781 & 0.298& $  20.96\pm   8.23$ & $   9.75\pm   2.04$ & $   2.18\pm   0.46$ &     1.3  & $5.2\times 5.2$  \\ 
     A1068$^{a}$ & 0.139& $  77.19\pm   1.86$ & $  14.51\pm   2.23$ & $   5.91\pm   0.91$ &    59.6  & $5.5\times 5.5$  \\ 
     A1423 & 0.213& $  46.02\pm   8.68$ & $   5.89\pm   1.69$ & $   1.69\pm   0.49$ &    3.1   & $3.9\times 10.3$ \\ 
     A1682 & 0.226& $  27.82\pm   8.79$ & $   5.87\pm   1.07$ & $   1.61\pm   0.29$ &     1.1  & $5.8\times 5.8$  \\ 
   A1758 N & 0.279& $  12.37\pm  12.17$ & $   4.82\pm   1.68$ & $   1.13\pm   0.40$ &     2.6  & -  \\ 
     A1763 & 0.223& $  18.38\pm   4.24$ & $   4.39\pm   0.55$ & $   1.22\pm   0.15$ &     2.0  & $1.4\times 3.4$  \\ 
     A2204 & 0.152& $  69.47\pm   3.19$ & $   9.08\pm   3.57$ & $   3.42\pm   1.34$ &   38.0   & $5.0\times 5.0$  \\ 
     A2219 & 0.228& $  20.55\pm   2.34$ & $   4.64\pm   0.29$ & $   1.26\pm   0.08$ &     5.3  & $4.6\times 4.6$  \\ 
     A2390 & 0.231& $  60.91\pm   1.16$ & $  10.81\pm   2.64$ & $   2.92\pm   0.71$ &    30.8  & -  \\ 
MS1455+2232 & 0.258& $  96.56\pm   1.10$ & $  19.62\pm   4.26$ & $   4.88\pm  1.06$ &    65.4  & -  \\ 
RXJ0439+0520 & 0.208& $  74.66\pm   6.20$ & $  11.81\pm   3.41$ & $   3.45\pm 1.00$ &   13.8   & -  \\ 
RXJ0439+0715 & 0.245& $  40.18\pm   3.51$ & $   5.45\pm   0.29$ & $   1.41\pm 0.08$ &     3.3  & -  \\ 
RXJ1720+2638 & 0.161& $  79.42\pm   2.77$ & $  12.64\pm   1.36$ & $   4.54\pm 0.49$ &   32.1   & amorphous source \\ 
RXJ2129+0005 & 0.234& $  72.78\pm   5.00$ & $   9.25\pm   2.86$ & $   2.47\pm 0.76$ &    8.5   & -  \\ 
  ZwCl1953 & 0.373& $  22.06\pm   8.33$ & $   4.48\pm   0.94$ & $   0.86\pm   0.18$ &     1.7  & $3.7\times 3.7$  \\ 
  ZwCl5247$^{b}$ & 0.229& $   <78.54$ & $   <8.83$ & $   <2.4$ &     0.6  & -  \\ 
\hline
\end{tabular}
\begin{quote}$^{a}$ The radio emission from A1068 is discussed in more
  detail in the text.
$^{b}$ The observation of ZwCl5247 is short for such a
  distant cluster.  It also does not have a cooling core, although it
  appears to have a central radio source, and so the X-ray emission is not
  strongly peaked at the centre.  The deprojection analysis results in
  a cooling time which is not well constrained at the centre.  We
  therefore give only the upper limits for the values obtained by the
  analysis for the quantities in this table.
\end{quote}
\end{table*}

\begin{figure}
\centering
\includegraphics[width=0.95\columnwidth]{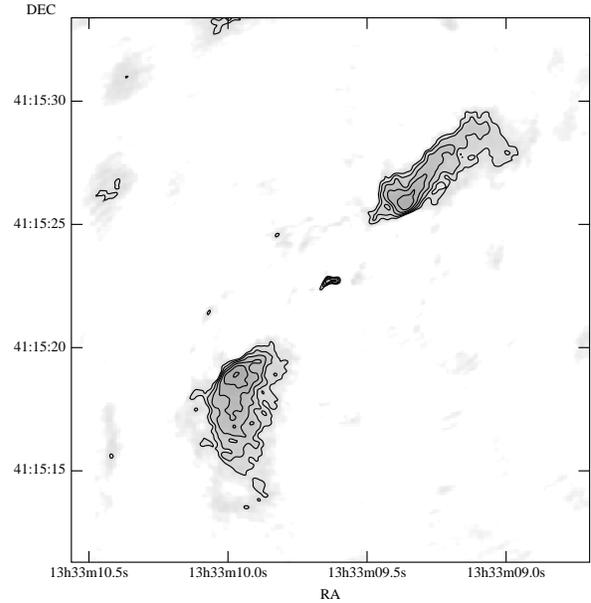}
\caption{\label{fig:a1763} The $8\ghz$ image of A1763 from the VLA in
  A-array.  This is the only cluster in which bubbles are expected but
  none seen in the X-ray emission, which also has a bi-lobed radio source at the centre where the structure of the
  radio emission was resolved in the archival VLA data.}
\end{figure}

Using the size of an expected bubble in these clusters, the X-ray
count total within the bubble was calculated from the current X-ray
observation.  This allowed the signal-to-noise of the observation to
be obtained\footnote{The signal-to-noise is calculated from the
  predicted size of the bubbles and the background subtracted counts per pixel within the
  central annular region.}.   The
difference in counts between the bubble centre and the rims
surrounding it was estimated from the bubbles in A2052 and Hydra A.
The counts in the bubbles from these two clusters are around 30 per cent lower than the counts
in the rims.  If bubbles are present, then they would need to be
detected above the noise at more than the $3\sigma$ level.  For this
to be the case, the signal-to-noise has to be greater than 13
(equivalent to noise at a 8 per cent level).  However, as the cluster
emission is changing over the scale of the bubbles the we may need a
higher signal-to-noise to be able to detect any depressions.  For a
$5\sigma$ detection, the signal-to-noise required is 36. This does require
knowing where the bubbles are so that the counts within the region
where a bubble is expected can be compared to that outside.  The results of this
analysis are shown in Table \ref{tab:hidden}.  

There are six clusters (A1068, A2204, A2390, MS1455+2232,
RXJ01439+0520 and RXJ1720+2638) where the central signal-to-noise of the observation is
such that if there are bubbles present within this region, then with the aid of any radio
emission to guide the eye, the bubbles should be detectable at the
$3\sigma$ level.  There are only three if the detection threshhold is
lifted to $5\sigma$ (A1068, A2204 \& MS1455), of which A1068 is
discussed below, A2204 is discussed in \citet{Dunn06f} and no high
resolution radio data was found for MS1455.  As this
signal-to-noise calculation has been calculated for the entire central
part of the cluster, the value obtained may over estimate the
signal-to-noise present at the location of the bubble if the core of
the cluster is very bright. 

The bi-lobed radio emission A1068 appears offset to the X-ray cluster
centroid. It would be expected that for a typical double source,
the lobes would sit either side of the X-ray peak (under the
assumption that the AGN is in the very centre of the cluster).  However, one ``lobe'' coincides
with the peak in the ICM emission, the other occuring at $38\kpc$ from
the centre.  In the vicinity of this more distant lobe, the signal to
noise is $\sim41$.  As no clear features are seen in the X-ray
emission, and the lobe locations are not typical, we investigated what
this radio source is coincident with.  Using the Sloan Digital Sky
Survey (SDSS, \citealt{Adelman07}) we extracted images of the sky centred on
the brightest cluster galaxy (BCG).  Aligning the radio and optical images
showed that the more distant (south-western) lobe is coincident with
another galaxy in the cluster ($z=0.136$ compared to $z=0.138$ for the
BCG).  The central ``lobe'' is therefore an
unresolved radio core.  This would also explain the lack of
bubble features in the high signal-to-noise X-ray emission, as no
bubbles are seen, given the lack of extended radio emission.

Using the Very Large Array (VLA\footnote{The
National Radio Astronomy Observatory is operated by Associated
Universities, Inc., under cooperative agreement with the National
Science Foundation.}) archive we attempted to find observations of
the clusters in which bubbles are expected to investigate the morphology of the central radio
source.  We analysed data at a variety of frequencies and array
configurations, depending what was available in the archive and which
had the longest time-on-source.

These clusters are all at redshifts higher than those in the B55
sample, and so we expected to resolve fewer of the central radio
sources.  Table \ref{tab:hidden} shows the expected bubble sizes and
measured radio source dimensions.  If no radio source dimensions are
given then the source was not resolved in the observations that were
analysed. 

The predicted bubble and observed radio source sizes are very similar
in A1068, A1763 (Fig. \ref{fig:a1763}) and A2204.  A1068 is discussed
earlier in this section and A2204 was discussed
in \citet{Dunn06f} and will not be covered further here.  However, the radio source size and the X-ray observation signal-to-noise
in A1763 are such that detecting any X-ray features of the bubble
would be impossible, even with a much longer exposure.  Under the
assumption that the natural course of events when an AGN ejects radio
emitting jets into the central regions of a cluster is that bubbles
are formed, then, as in the B55 cluster sample \citet{Dunn06f}, we are
missing a large number of bubbles.  This may be a timescale effect,
currently in some of these clusters the AGN is not inflating bubbles,
or they are present but too small to be detected with either the
current X-ray or radio data.

In nearby clusters where X-ray and radio spatial resolution do not
pose a problem for studying the morphology of the AGN and cluster,
whenever there is a lobed radio source, there are always signs of an
interaction between the relativistic and thermal plasmas.  Therefore,
for more distant objects, where even the spatial resolution of
{\it Chandra} is insufficient to allow the detailed investigation of
the interaction, perhaps the radio emission alone will be enough to
determine the action of the AGN on the cluster.  If the two plasmas do
not significantly mix over a timescale of $\sim 10^8 \yr$, then if the
AGN is actively producing jets, bubbles would be the natural
consequence, along with the associated heating of the ICM.  Therefore,
if the bubbles do allow a good calibration of the AGN energy injection
rate, the dimensions of the radio emission rather than of any X-ray
features could be used to calculate the heating from the AGN.

\section{Clusters requiring heating}\label{sec:heating}

All the clusters which require heating (but do not have clear bubbles
-- either reported or determined from the {\it Chandra} data) harbour
a central radio source (as determined from the NVSS).  Their expected
bubble sizes have been calculated in Section \ref{sec:radio}.

\begin{figure*}
\centering
\includegraphics[width=0.3\textwidth]{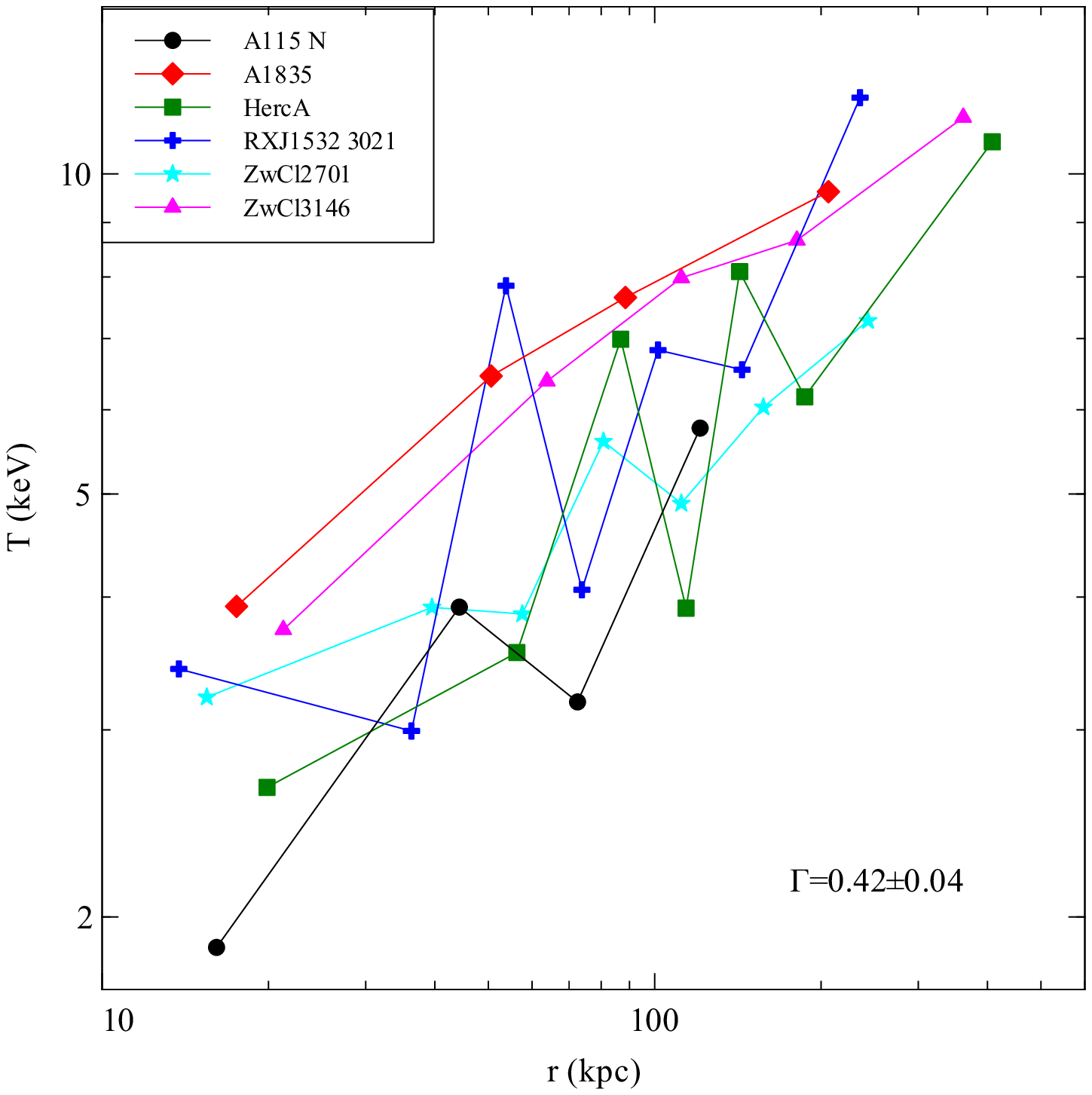}
\includegraphics[width=0.3\textwidth]{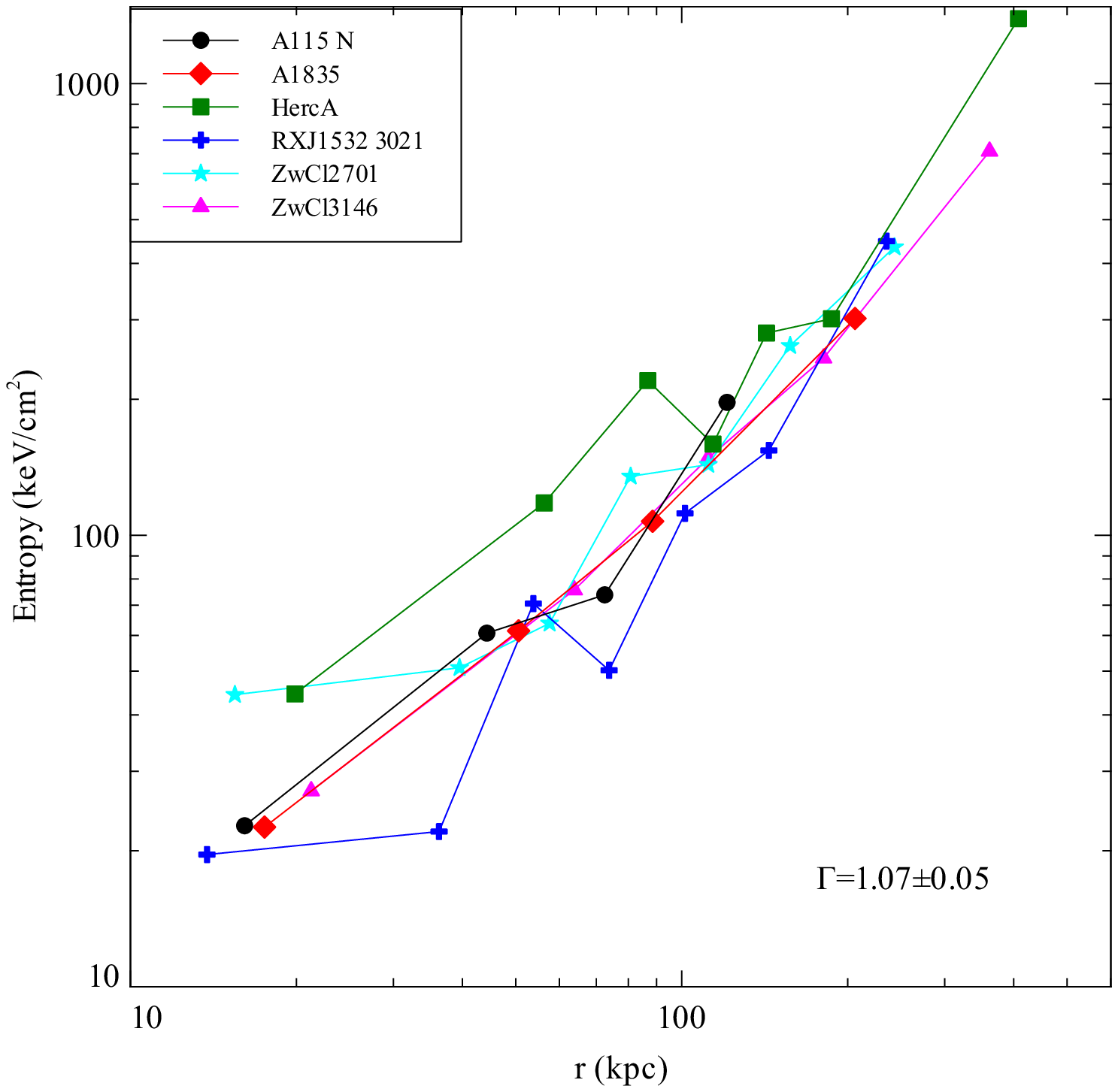}
\includegraphics[width=0.3\textwidth]{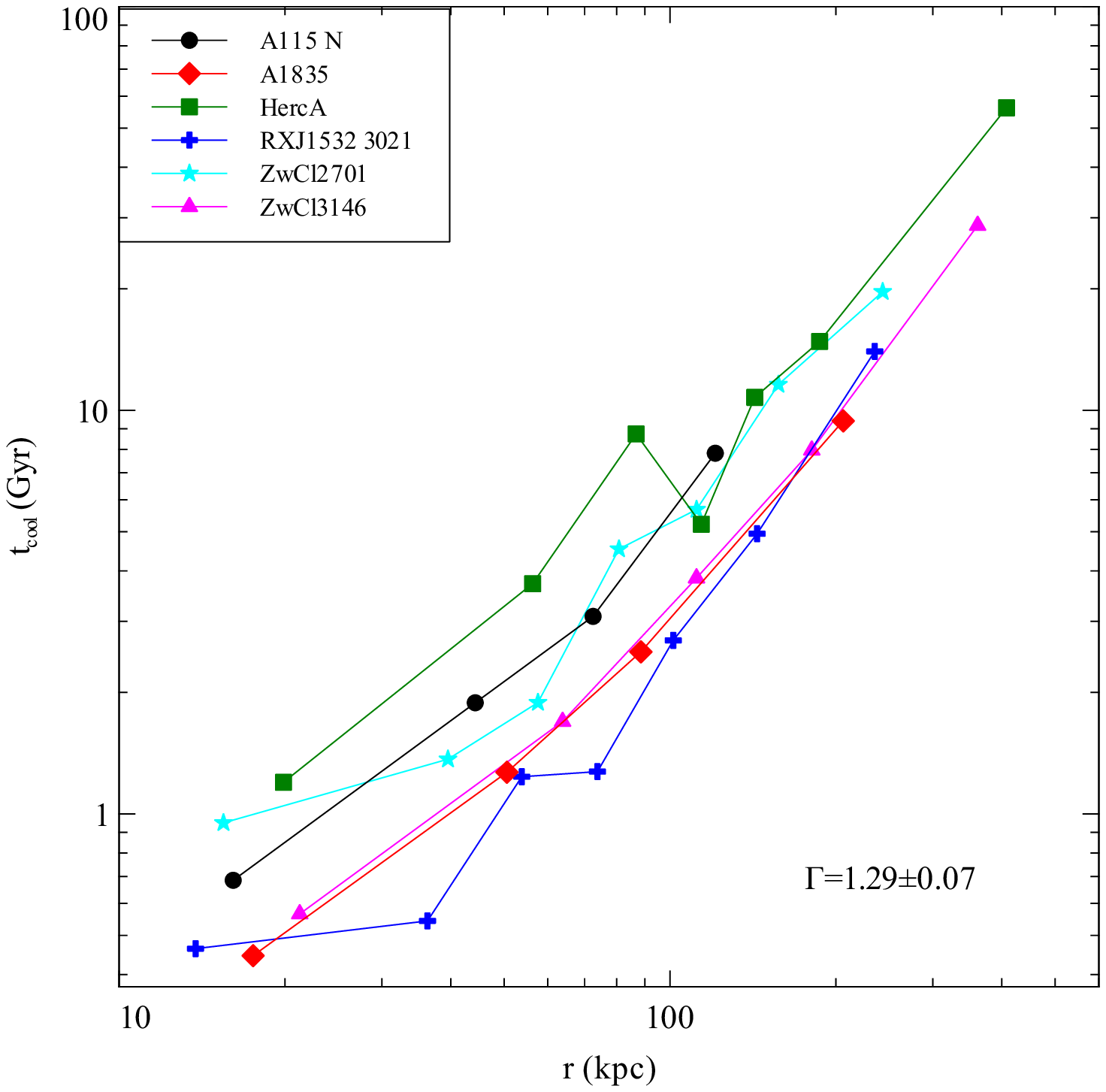}
\includegraphics[width=0.3\textwidth]{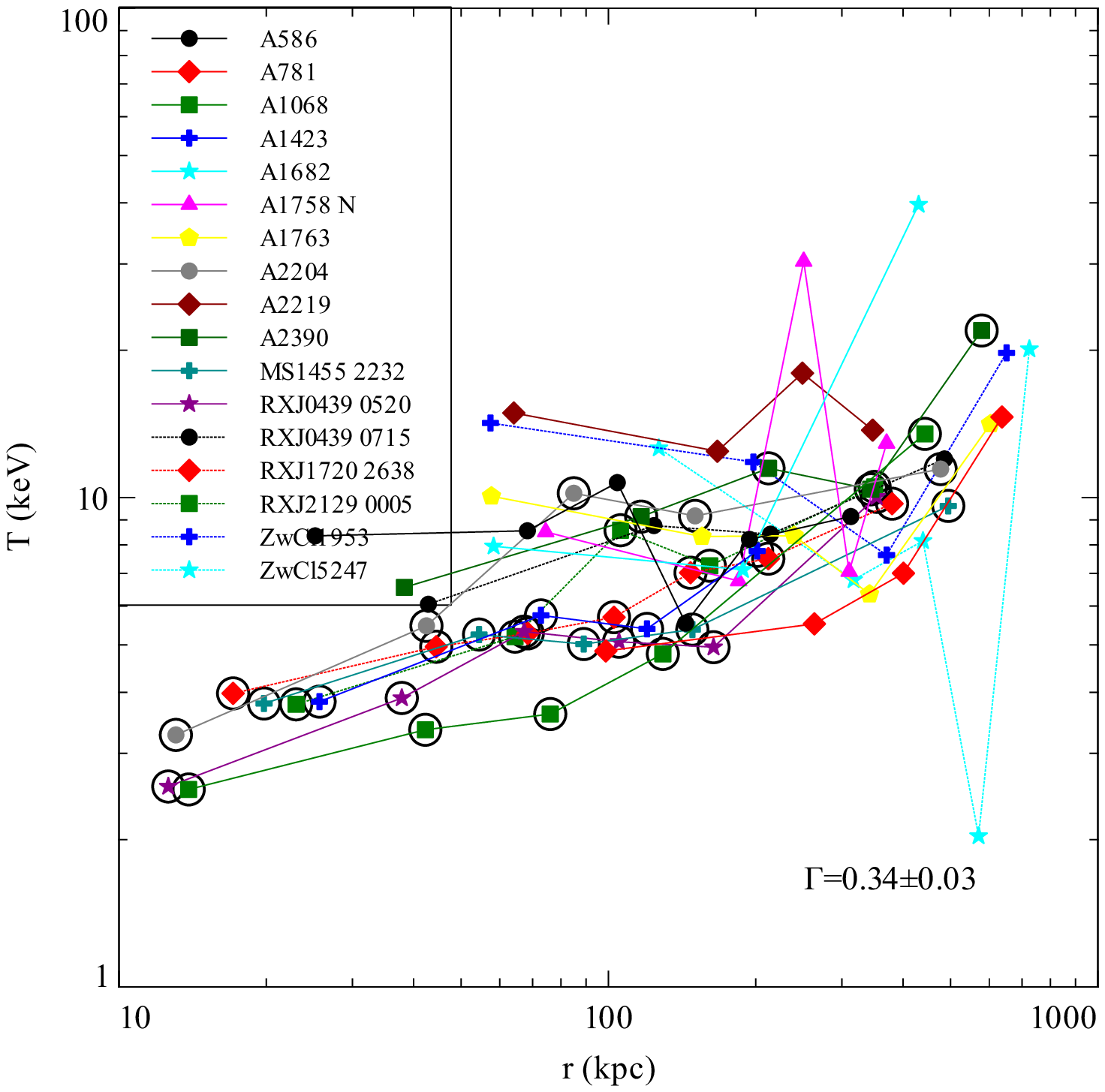}
\includegraphics[width=0.3\textwidth]{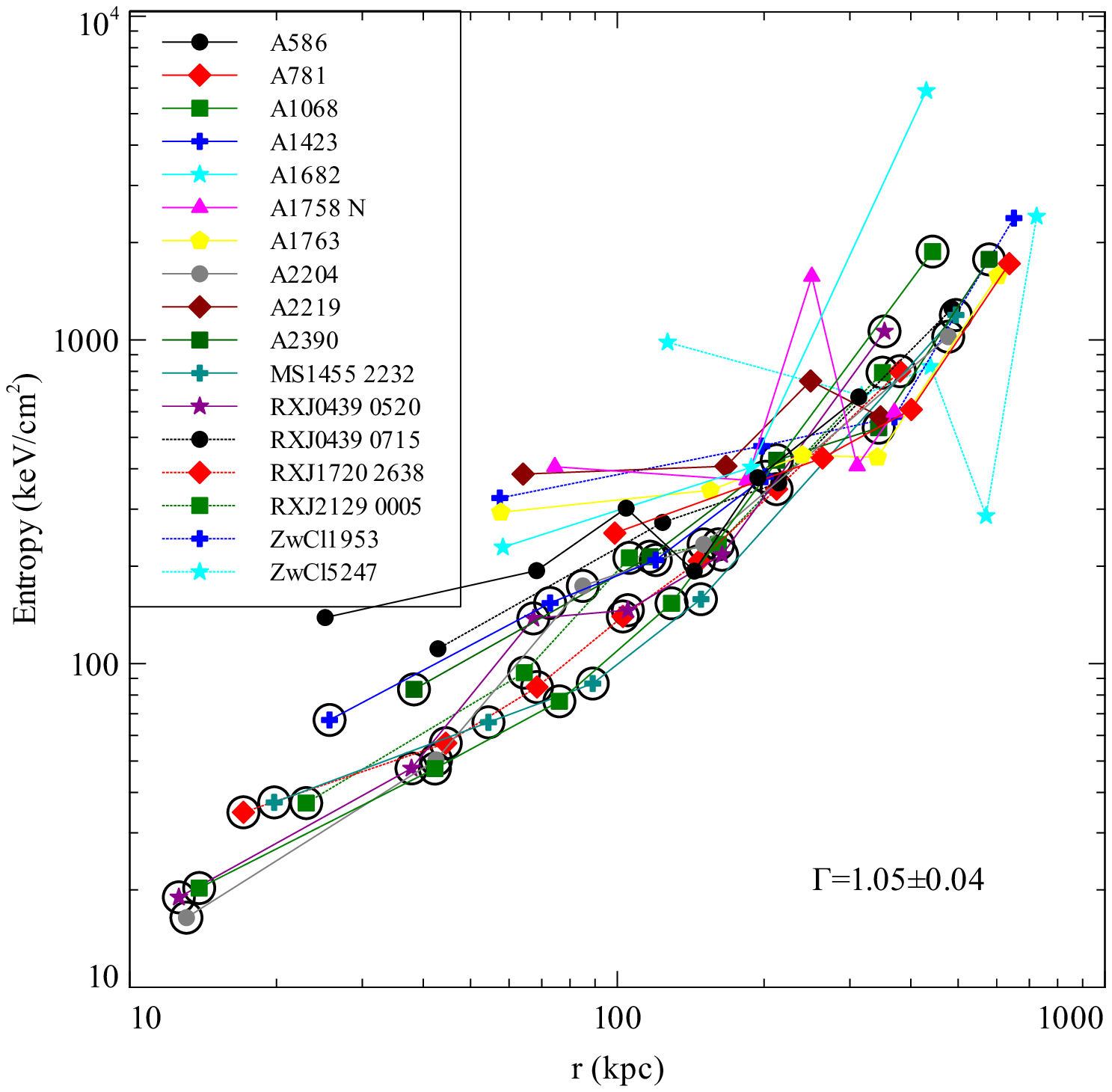}
\includegraphics[width=0.3\textwidth]{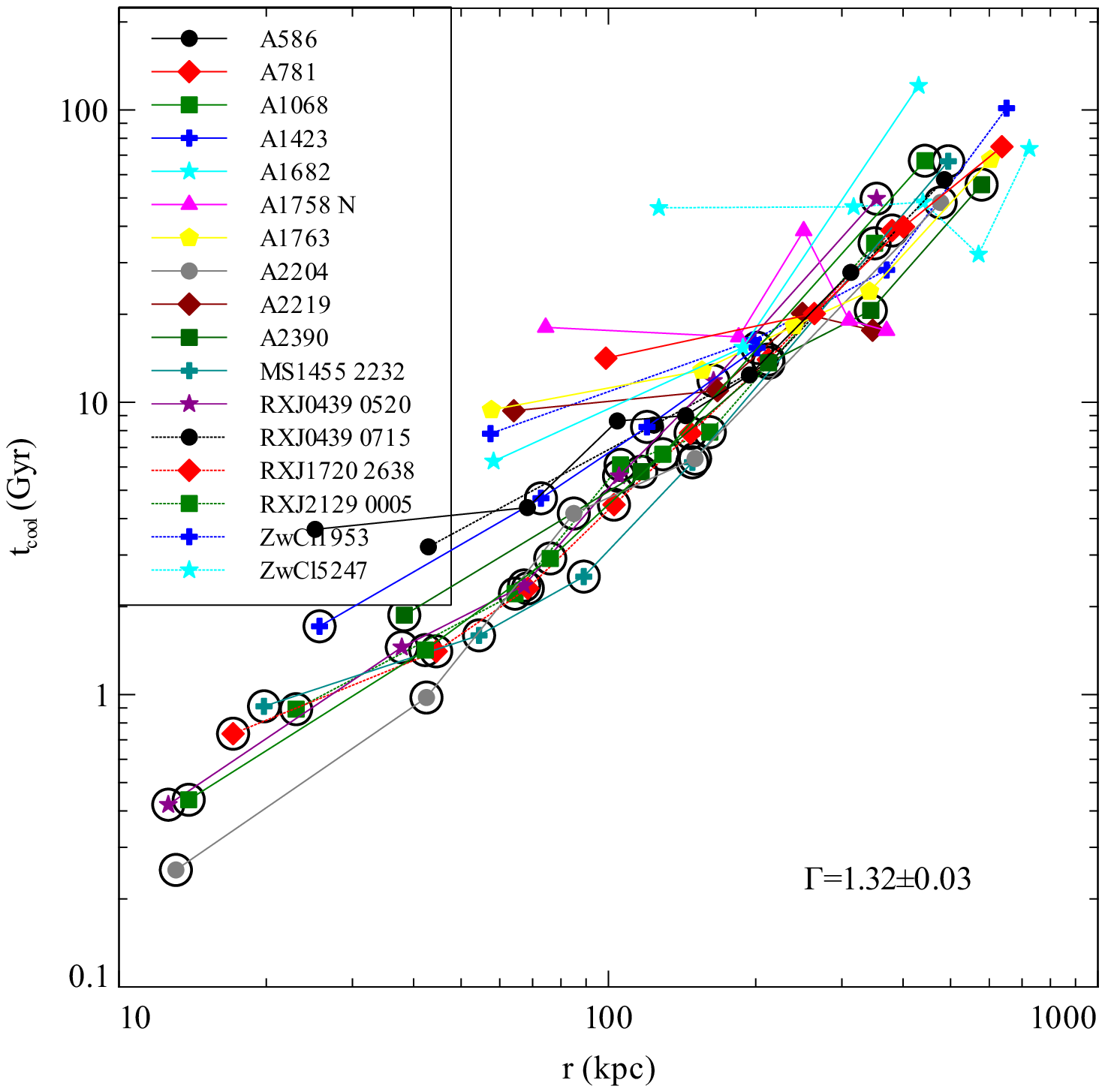}
\includegraphics[width=0.3\textwidth]{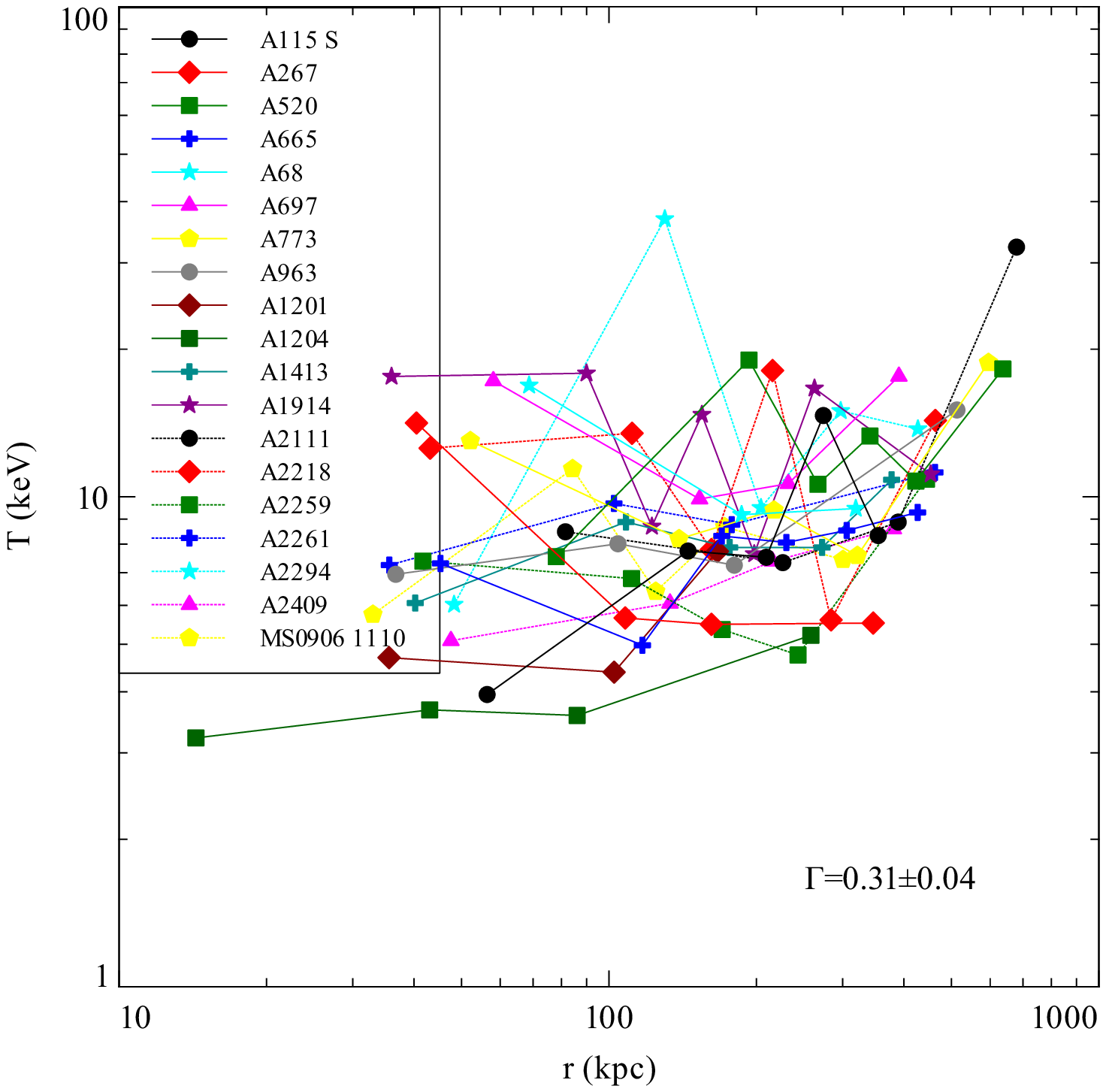}
\includegraphics[width=0.3\textwidth]{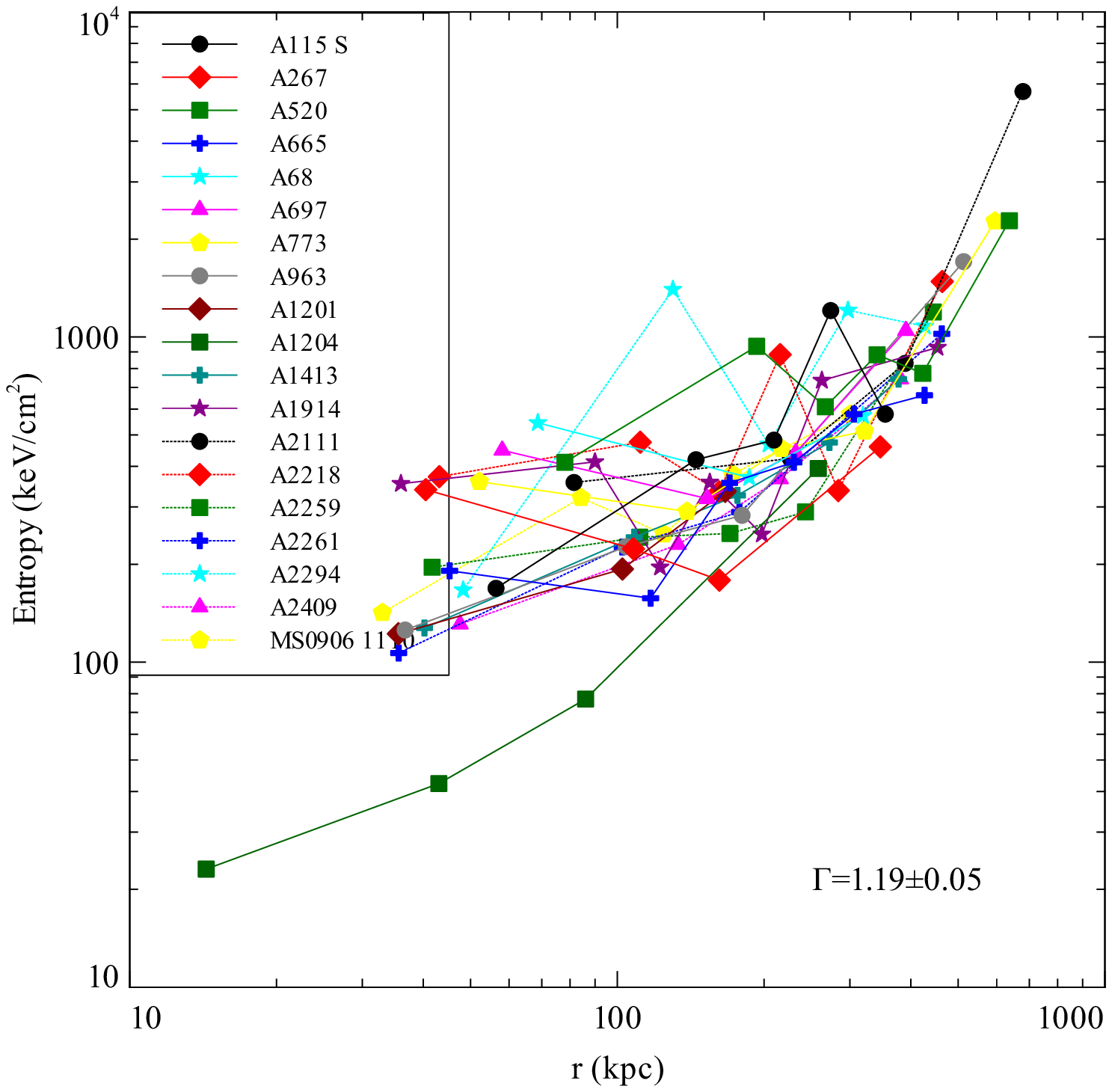}
\includegraphics[width=0.3\textwidth]{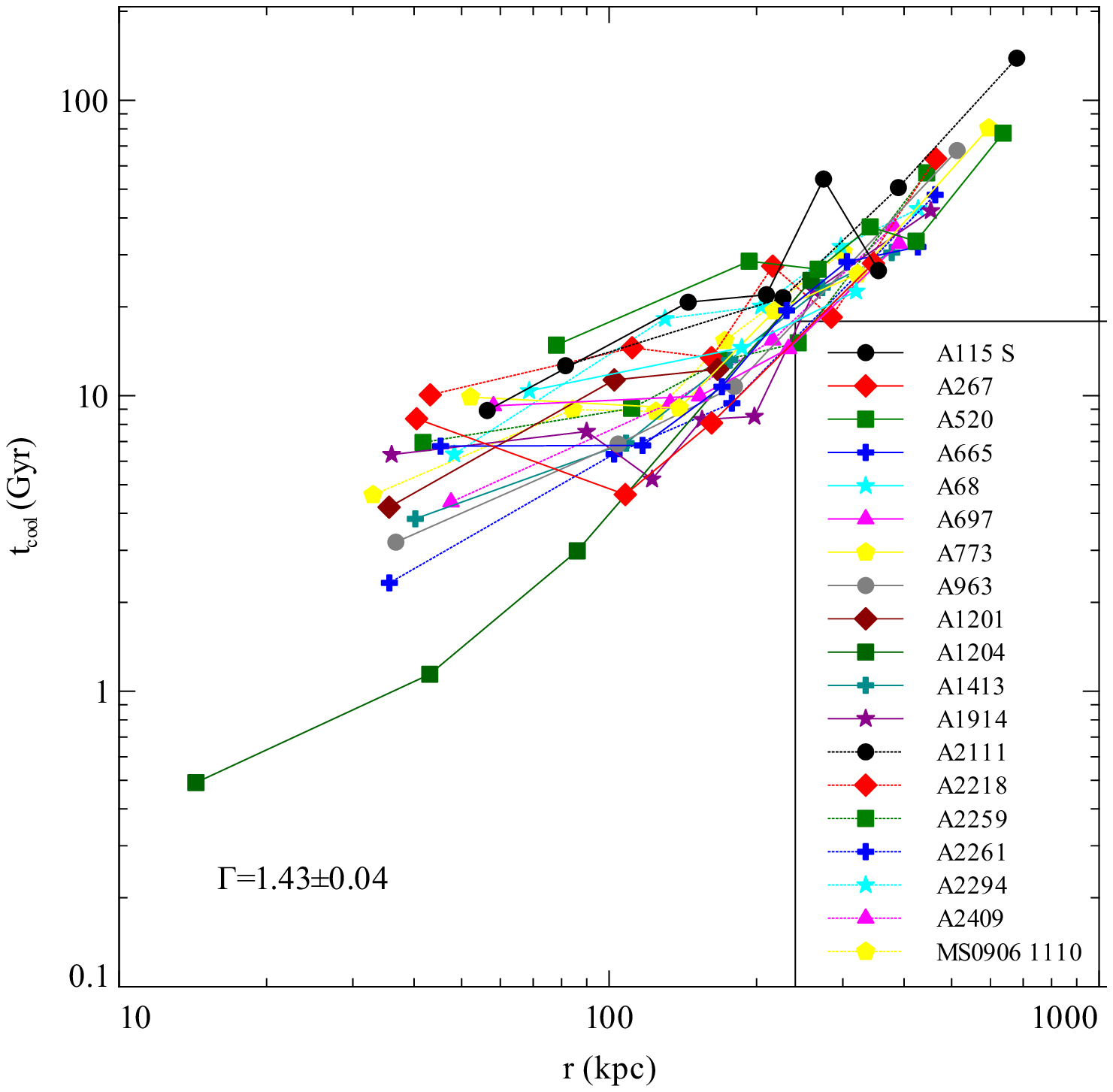}
\caption{\label{fig:profiles} {\scshape left}: Temperature, {\scshape
    centre}: Entropy and {\scshape right}: Cooling time profiles for
    {\scshape top}: Clusters with clear bubbles, {\scshape middle}:
    Clusters with central radio sources and {\scshape bottom}:
    clusters with no central radio source.  Best fitting powerlaw
    slopes, $(\Gamma)$,  to all the date on the plots are shown.  The clusters which
    require heating are highlighted by a large circle in the plots for the clusters
    harbouring a central radio source.}
\end{figure*}

As a comparison between the three different sub-samples, Fig. \ref{fig:profiles}
shows the temperature, entropy and cooling time profiles for the
different sub-samples.  There are only six clusters in the sub-sample
containing clear bubbles and so small number statistics are relevant.  Fig. \ref{fig:profiles} also shows the best fitting powerlaw
slope to all the data points.  It is interesting to note that in the
average slope for all
three parameters, the slope for the clusters with radio emission is always
in between those with bubbles and those with neither.  The temperature
slopes of the radio and no-radio clusters are almost identical,
whereas with the entropy and cooling time, it is the bubbled and radio
clusters which have very similar slopes.

The clusters which require some form of heating are highlighted in
Fig. \ref{fig:profiles}.  The clusters which require
some form of heating do appear to be the coolest clusters and those which
have the lowest central entropy.  They are naturally those with the
shortest central cooling time as this was one of the selection
criteria.  As is clear from Fig. \ref{fig:profiles}, these clusters are those in
which the deprojection resulted in the smallest central bin.  As
discussed in \citet{Bauer05}, the radius of the innermost annulus
depends on the cluster distance and the signal-to-noise of the
observation.  The latter parameter is related to whether the cluster
exhibits a cool core or not, as cool core clusters have higher central
densities and hence have brighter X-ray cores. 

There is a further possible bias towards having detailed/deep observations
of cool core clusters and less deep observations of other clusters.  A
quick calculation of the average exposure for clusters which have
bubbles, a radio source or neither results in $33.5$, $28.4$ and
$24.3\kilos$ respectively.  Although these exposure times are not
significantly different (standard deviation $\ge 17$) it is interesting to note
that the clusters which are of most interest to the heating and
cooling balance are those which appear to have the longest exposure
times.  These clusters, as they are cool core, are also the
brightest.  For a balanced analysis the non-cool core clusters
probably should
have a longer average exposure times as they are less bright.

If the entropy data are split into those clusters which
require some form of heating and those that do not, then there is a large
difference in the slopes of the average entropy profiles.  The different best
fitting slopes are 1.01 and 0.72 respectively.  Ignoring A1204 in the
entropy fits for the clusters with no radio data results in a slope of
0.77.  A1204 only just fails on the temperature drop selection
criterion for requiring heating.  

The average cooling time profiles of the clusters with bubbles and
those with a radio source match the profiles presented in
\citet{Voigt04} of $t_{\rm cool}\propto r^{1.3}$.

\section{Discussion}\label{sec:discuss}

\subsection{Combining the Samples}\label{sec:discuss:comb}

In order to study a large number of clusters over a range of redshifts
we combine the BCS and the B55 samples.  As A2204 is present in both
we have used the central cooling time from the analysis presented in
this manuscript in the following discussion.  We therefore have $30+42-1=71$
clusters in the combined sample.

The B55 and BCS samples have slightly different selection criteria.
The BCS sample selects clusters with $\delta>0^\circ$, $|b|\geq 20^\circ$, $z<0.3$ and
$S_{0.1-2.4\kev}>4.4\times 10^{-12}\ergpcmsqps$, with the extended
sample down to $S_{0.1-2.4\kev}>2.8\times 10^{-12}\ergpcmsqps$.  The
B55 sample has $S_{2-10\kev}>1.7\times 10^{-11}\ergpcmsqps$ as its
criteria and includes 9 clusters at low galactic latitudes $(|b|<
20^\circ)$.  Both are around 90 per cent flux complete.  We do not
adjust the selection criteria for consistency between the two
samples. 

So that both the BCS and B55 cluster samples have the same data
reduction and spectral fitting applied we have re-reduced the B55
cluster data.  This has allowed us to take advantage of the improved
data-reduction script -- e.g. taking into account the chip gaps in
ACIS-I observations.

\subsubsection{Cooling times}\label{sec:discuss:comb:tcool}

\begin{figure}
\centering
\includegraphics[width=0.95\columnwidth]{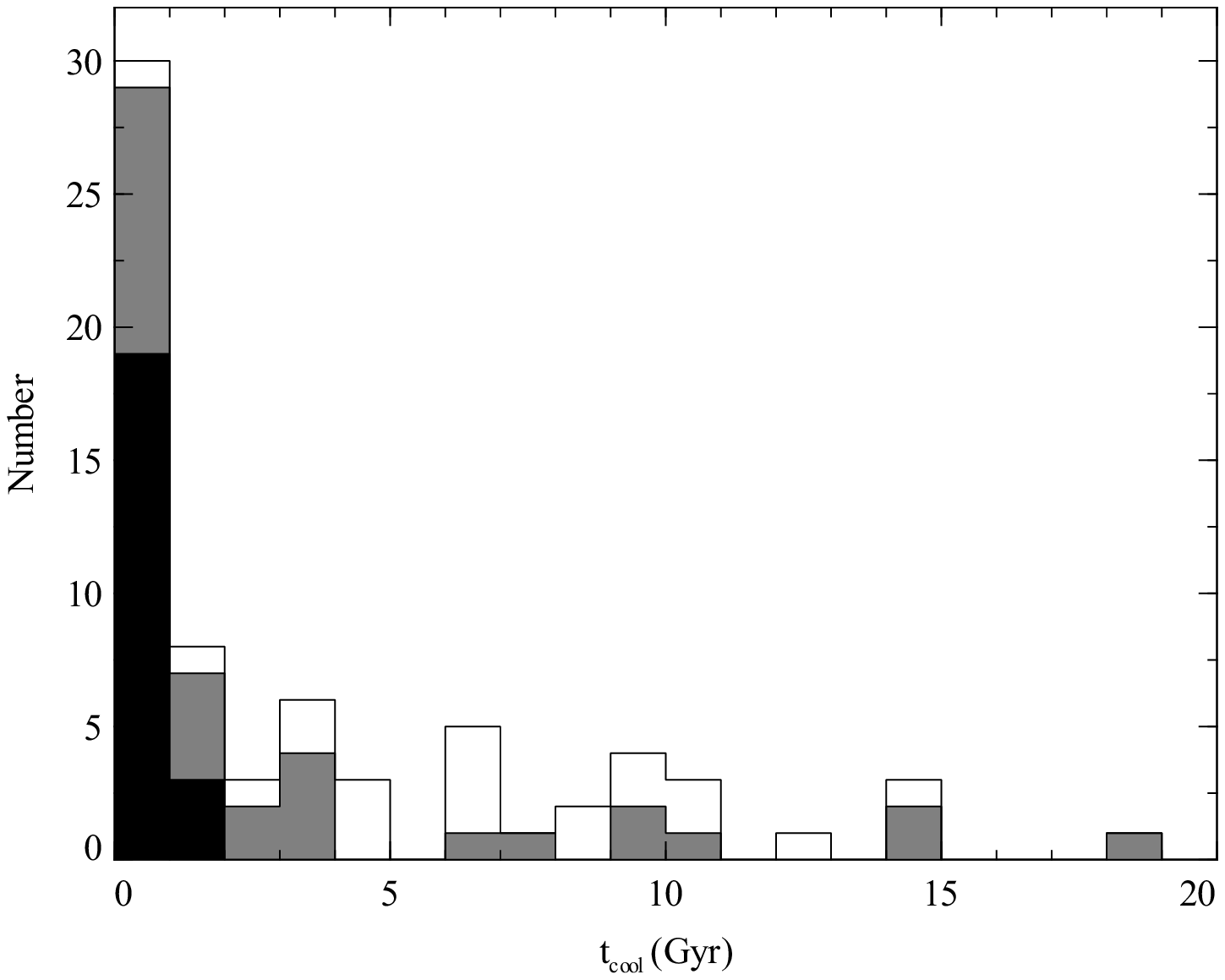}
\includegraphics[width=0.95\columnwidth]{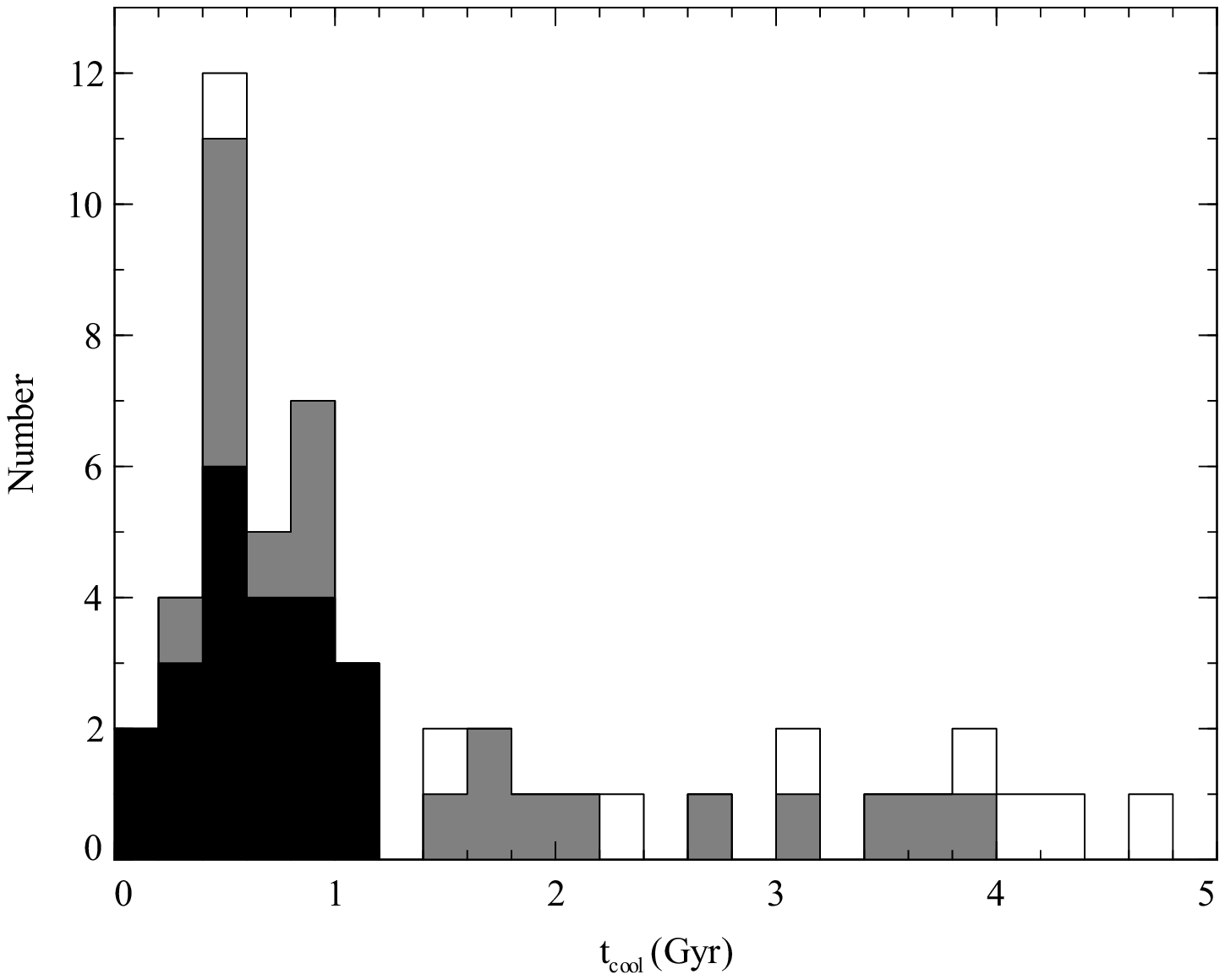}
\caption{\label{fig:tcool_comb}{\scshape top}: The distribution of $t_{\rm cool}$ in the
  combined cluster
sample. {\scshape bottom}: A finer binning of the clusters with a
  short central $t_{\rm cool}$.  The black bars indicate those clusters which harbour clear
bubbles, and the grey ones, those with a central radio source.}
\end{figure}

Fig. \ref{fig:tcool_comb} shows the distribution of cluster
central cooling times.  Rather than take the values from
the analysis of {\it ROSAT} data \citet{Peres98}, we have repeated the
extraction of the central cooling time performed here on the BCS
sample, on the B55 sample.  We note that not all of the clusters in the high redshift
end of the BCS sample have been observed with {\it Chandra}; in
\citet{Dunn06f} clusters without short central cooling times or
central radio sources were discarded from further analysis, whereas
all clusters for which {\it Chandra} observations exist have been
analysed in this work.  As a result the fraction of clusters
harbouring bubbles/radio sources as a function of the central cooling
time are estimates.

What can be seen from Fig. \ref{fig:tcool_comb}, however, is that
radio sources (with or without clear bubbles) are much more common in
clusters with central cooling times of $4\gyr$ or less.  In fact there
are only 10/24 clusters with $t_{\rm cool}\geq 4\gyr$ which have central radio sources.  In the
distribution with finer binning, the peak in the central cooling times
appears at around $0.6\gyr$.  In the current combined samples, there
appear to be no clusters which harbour
clear bubbles and have central cooling times $>1.2\gyr$.  Very few
clusters (5/47) with central cooling times shorter than $4\gyr$ have no
evidence for a central radio source (from the NVSS).

Although the two samples combined here have not been created in the
same way (lack of non-cool core/non-radio source harbouring clusters
in the B55 analysis) there appears to be an excess of clusters which
have a cooling time of $\lesssim 1\gyr$.  To take into account the
effect of all the clusters in
the B55 sample which have not yet been analysed on Fig
\ref{fig:tcool_comb}, we look at the distribution of $t_{\rm
  cool}$ at $>3\gyr$.  The distribution is fairly flat, so even if all the
remaining clusters follow this flat distribution, then the ``pile-up'' of
clusters at $t_{\rm  cool}<1.2\gyr$ would still be seen.

An explanation of this pile-up could result from the feedback cycle
suggested as a solution to the cooling flow problem.  Only once the
central cooling time is less than $\sim 1 \gyr$ will the AGN be active
and inject energy into the central regions of the cluster and create
the tell-tale bubbles.  It would depend on whether the AGN activity is
gentle and relatively continuous or explosive and intermittent as to
what the final state of the cluster would be.  An explosive and
energetic AGN outburst would increase the ICM temperature, and hence
cooling time by a large amount.  As a result the next injection event
would occur after a long dormant period.  However, a gentle injection of energy
would only reheat the ICM a small amount, and so the next injection
event would take place relatively shortly thereafter.

The large number of clusters at short cooling times, some of which
have radio sources but no clear bubbles, implies that the majority of
injection events are the ``gentle and often'' kind; as most clusters
with radio sources (dormant AGN?) and bubbles are found with short
central cooling times.

\citet{Bauer05} investigated the cooling time profiles of the {\it
  Chandra}-observed BCS clusters.  There are five more clusters in our
sample as we use a lower minimum redshift for our sample selection and
there have been further observations of BCS clusters.  We compare our central cooling times
to theirs and find that on average ours are shorter by around a factor
0.7.   There are some differences between the data-reduction methods
of the two studies.  We do use the cluster centroid as calculated by the {\scshape ciao}
software as in \citet{Bauer05}, but then visually inspect the annular regions and manually
change their location to match the surface brightness peak (12
clusters).  We also allow the galactic absorption to be unconstrained
in the spectral fitting.  \citet{Bauer05} state that using only the
centroid rather than the surface brightness peak could bias their
central cooling times up by up to a factor 3.  

\subsubsection{Heating-Cooling
  balance}\label{sec:discuss:comb:balance}

\begin{figure}
\centering
\includegraphics[width=0.95\columnwidth]{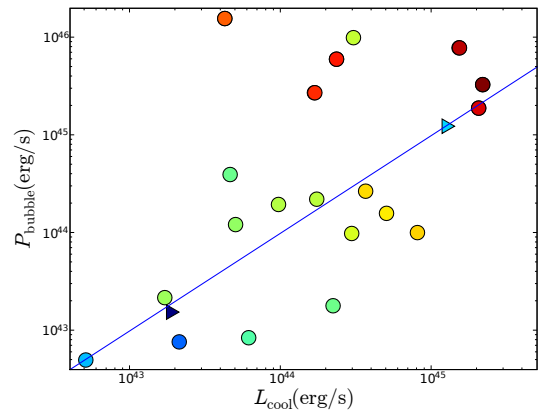}
\caption{\label{fig:LcoolPbubble} The rate of energy loss via X-ray emission from within
  the cooling radius of the cluster versus the bubble power from the
  AGN.  These are for the clusters which harbour clear bubbles.  The
  two triangle symbols are for M87 and Perseus where $L_{\rm cool}$
  could not be determined from a single ACIS chip.  The colour scale
  is the logarithm of the redshift, blue is low redshift, red is high
  redshift.  The line is for exact balance between $P_{\rm bubble}$
  and $L_{\rm cool}$.}
\end{figure}

We add the six clusters in the BCS sample which harbour clear bubbles
to those 16 from the B55 sample and update Fig. 2 of \citet{Dunn06f} for
the cooling luminosity and the bubble power (Fig. \ref{fig:LcoolPbubble}).  The colour scale shows
the redshift of the cluster.  The six new clusters are all at the top
of the plot in red as they are all at $z>0.15$.  

The average distance, as a fraction of the cooling radius,  out to which the bubbles can offset the X-ray
cooling from the combined sample is
$r_{\rm heat}/r_{\rm cool}=0.92\pm0.11$.  The fraction of the
X-ray cooling within the cooling radius offset by the action of the
AGN is $  0.90\pm0.13$.  We ignore those clusters
where the bubbles are sufficiently powerful to offset all the cooling
within the analysed region, and so only a lower limit on the distance
is obtained.  Including these clusters gives $r_{\rm heat}/r_{\rm
  cool}=2.10\pm0.40$ and the fraction of $20.92\pm15.11$.  Although
there are selection effects, especially at the high redshift end of
the sample, we can conclude that on average the bubbles counteract the
X-ray cooling within the cooling radius.

The problem with this plot is that it shows an instantaneous measure of
$L_{\rm cool}$ and $P_{\rm bubble}$ rather than a time average.
However for the lower redshift sources it is still reasonable to say
that on average the bubble power is sufficient to counteract the X-ray
cooling.  At the higher redshifts all clusters lie on or above the
line of exact balance.  Under the assumption that clusters at
redshifts of around $0.2-0.3$ are similar to those in the local
Universe, we would expect to see some clusters which have bubbles
which appear to be insufficient to counteract all the X-ray cooling.

\begin{figure}
\centering
\includegraphics[width=0.95\columnwidth]{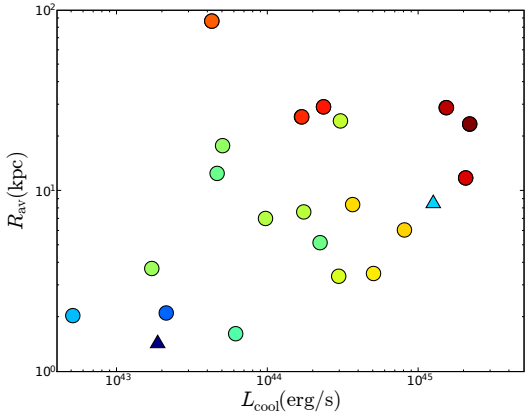}
\caption{\label{fig:LcoolRav} The rate of energy loss via X-ray emission from within
  the cooling radius of the cluster versus the average bubble size for
  the cluster.  These are for the clusters which harbour clear bubbles.  The
  two triangle symbols are for M87 and Perseus where $L_{\rm cool}$
  could not be determined from a single ACIS chip.  The colour scale
  is the logarithm of the redshift, blue is low redshift, red is high
  redshift.}
\end{figure}

Fig. \ref{fig:LcoolRav} shows that there is a trend between the average size of
a bubble and the X-ray cooling luminosity.  The smaller $L_{\rm cool}$,
the smaller the detected bubbles.  This trend does point to some
form of feedback between the X-ray cooling and the kinetic luminosity
of the central AGN.  

In Fig. \ref{fig:LcoolPbubble} there is also a general trend in the redshift of the cluster as the
bubble power increases, from bottom left to top right.  We do not see
any high-redshift, low-power bubbles, which is not unexpected.  At high
redshifts these bubbles would be too small to detect with the spatial
resolution of {\it Chandra}.  We do, however, see some low-redshift
clusters with energetic outbursts -- Cygnus A and Perseus.  This is
also demonstrated in Fig. \ref{fig:LcoolRav}.  There are few large
bubbles at low redshifts, as a result of the small sampling volume;
whereas at high redshifts there are few small bubbles as they would
not clearly be resolved.

Figs. \ref{fig:LcoolPbubble} and \ref{fig:LcoolRav} imply that we are missing a number
of bubbles, mainly at high redshift, that have not as
yet been detected.  As they have not as yet been seen by {\it
  Chandra}, then other methods for their detection will have to be
developed if their effect on the clusters is going to be analysed in
detail, as {\it XEUS}\footnote{X-ray Evolving Universe Spectrometer,
  \url{http://www.rssd.esa.int/index.php?project=XEUS}} will have at
best $2\,$arcsec spatial resolution.  As can be seen in
Fig. \ref{fig:tcool_comb}, most clusters which have short cooling
times harbour radio sources, and some have clear bubbles.  It could be
argued that the natural consequence of an AGN at the centre of a
cluster is that, if it is currently producing jets, it produces
bubbles and injects energy into the centre of the cluster.  The
GHz radio emission could then be used as a tracer of the current
size of the bubbles, and from this the energy input of the AGN be
quantified. 

Another interpretation is that the AGN in high redshift clusters are
undergoing their first and hence explosive outburst, and that they
will, over time, head towards a steady balance between $L_{\rm cool}$
and $P_{\rm bubble}$.  However, as Cygnus A and Perseus appear in very
similar locations to the high redshift clusters and Perseus has clear
evidence for previous outbursts then this explanation is a little uncertain.

\begin{figure}
\centering
\includegraphics[width=0.95\columnwidth]{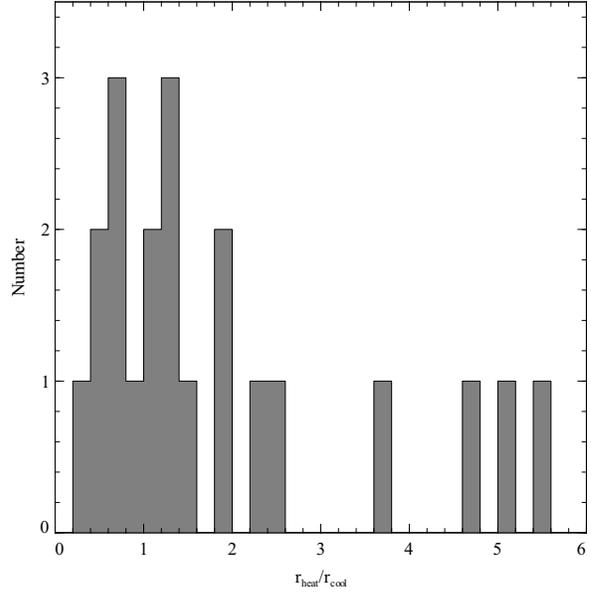}
\caption{\label{fig:r_rcool} The distribution of radii (as a
  fraction of the cooling radius) out to which the energy in the
  bubbles can offset the X-ray cooling for all clusters in the
  combined sample which have clear bubbles.  Hercules A is off the
  right hand side of this figure ($r_{\rm heat}/r_{\rm cool}=13.1$).}
\end{figure}

The distribution of radii out to which the energy in the bubbles
can offset the X-ray cooling (see Fig. \ref{fig:r_rcool}) show that a
large number of clusters occur around equality with a tail towards
``overheating.''  As most clusters are close to $r_{\rm heat}/r_{\rm
  cool}=1$ it is reasonable to assume that even out to $z\sim0.4$
there is a balance between AGN heating and X-ray cooling, though the
caveat of missing small bubbles at high redshifts should be noted.

\subsubsection{Duty Cycle}\label{sec:discuss:comb:duty}

A significant fraction of the clusters in the combined sample have a
large heating effect on the cluster.  There are two clusters in the
B55 sample which have $r_{\rm heat}/r_{\rm cool}>2$ (A2052 \& Cygnus
A) and there are four in the BCS sample (A115, Hercules A, ZwCl2701
\& ZwCl3146).  As these outbursts are going to arise in clusters in
which some form of heating is expected, the parent sample of these is
34 (20 from B55 and 14 from BCS).  Therefore 18 per cent
(6/34) of clusters in this sample have extreme outbursts, and hence the duty cycle is
also 18 per cent.

\subsubsection{Bubble Sizes}\label{sec:discuss:comb:sizes}

\begin{figure}
\centering
\includegraphics[width=0.95\columnwidth]{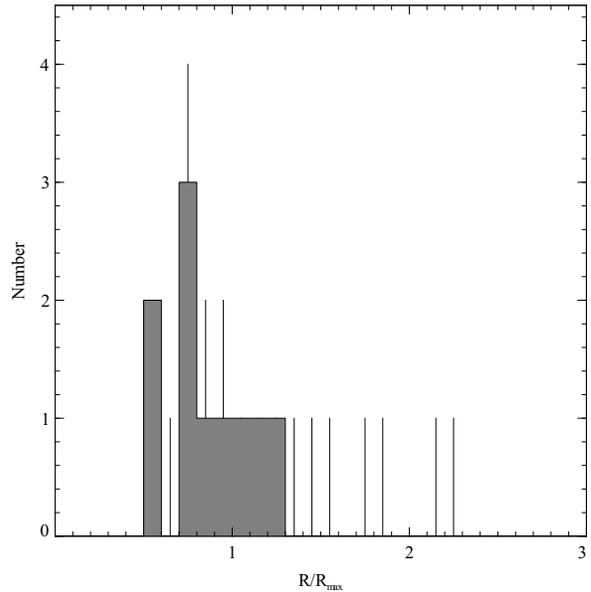}
\caption{\label{fig:RRmax} The distribution of the ratio of the
  average to the ``maximum'' size of the bubbles.  The uncertainties
  are estimated using a Monte Carlo simulation of the calculations.}
\end{figure}

We repeat the analysis in \citet{Dunn06f} of comparing the observed
size of the bubbles to the size expected if they offset all the X-ray
cooling within $r_{\rm cool}$.  For a detailed description of the
method see Section 6.1 of \citet{Dunn06f}.  The ``maximum'' radius of
the bubble is defined as
\begin{equation}
R_{\rm max}=\sqrt{\frac{3\mathcal{C}(L_{\rm cool}/2)}{16\pi p_{\rm
      th}v_{\rm c_s}}},
\end{equation}
and is compared to the average radius of the bubble $R_{\rm
  av}=\sqrt[3]{R_{\rm l}R_{\rm w}^2}$.  We assume that an AGN will
produce two bubbles, hence each will offset $L_{\rm cool}/2$.  We use Monte Carlo simulations
of the data to allow estimates to be placed on the uncertainties in
the distribution\footnote{The values of $R_{\rm av}/R_{\rm max}$ were
  binned $10^4$ times using a Gaussian distribution for the input
  uncertainties.  The interquartile range in the values for the bins
  is shown by the error bars in Fig. \ref{fig:RRmax}}.  Only clusters with young bubbles were used for this
calculation.  The distribution is shown in Fig.  \ref{fig:RRmax} and
still shows a peak at around $R_{\rm av}/R_{\rm max}\sim 0.7$.

The peak has not moved significantly from that found by
\citet{Dunn06f}, which is not surprising as only 3 clusters (6
bubbles) have
been added.  The explanations for this distribution as outlined in
\citet{Dunn06f} are therefore all still valid.  The bubbles with a
small value of $R_{\rm av}/R_{\rm max}$ are small for their host
cluster, and so are likely to still be growing.  Therefore, it is
natural to state that all the bubbles with $R_{\rm av}/R_{\rm max}>1$
are large, i.e. have detached from the core of the cluster and are
expanding in the reduced pressure of the higher cluster atmosphere.
Some of these large bubbles may involve AGN interactions which are not well
described by bubbles, e.g. Hydra A \citep{Nulsen04}.

As bubbles spend a larger fraction of their lifetime close to their full
size rather than growing with small radii, it would be expected to see
the majority of clusters with $R_{\rm av}/R_{\rm max}\sim 1$.  To
shift the peak in the distribution from 0.7 to 1.0 requires $R_{\rm
  max}$ to reduce.  This could be achieved by reducing $L_{\rm cool}$,
but this is an unlikely solution.  We have already taken a fairly
extreme definition for the cooling radius, and it would be more
sensible to increase the cooling radius to move it closer to other
definitions of cool core clusters.

\subsection{Evolution in the Cluster
  Samples}\label{sec:discuss:z_distn}

Comparing the number of clusters requiring heating/harbouring radio
sources between the BCS and B55 samples show that they are very
similar.  The fraction of clusters harbouring radio sources is almost
exactly the same for each sample.  The fraction of clusters requiring
heating is slightly larger for the BCS sample (33 versus 25 per
cent).  This may be a bias towards {\it Chandra} observations of
cooling clusters, but may also be a true increase in the fraction of
clusters requiring heating.

By combining the clusters in the BCS and B55 samples we have two well
defined samples spanning a redshift range of $0.0\leq z \leq 0.37$.
However the treatment of the two samples has been different, and so
when investigating any changes in the clusters with redshifts we have
to redefine the samples so that selection effects do not influence our
results.

\citet{Dunn06f} did no analysis of clusters in the B55 sample which
had no bubbles or central radio source, whereas in the analysis
presented here, all clusters which had {\it Chandra} observations were
studied.  To minimise any selection effect we concentrate on clusters
with a radio source and/or bubbles.

\begin{figure*}
\centering
\includegraphics[width=0.3\textwidth]{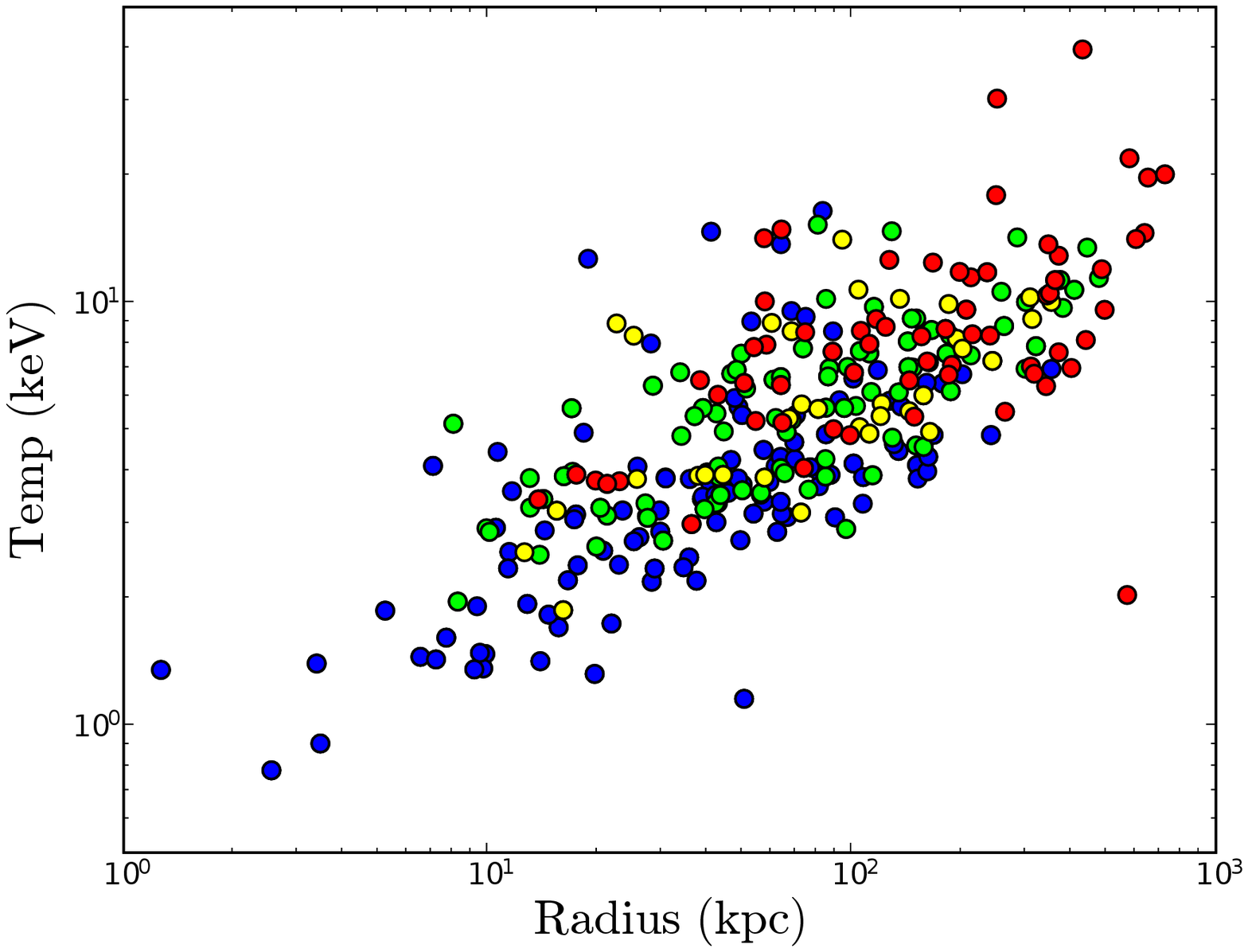}
\includegraphics[width=0.3\textwidth]{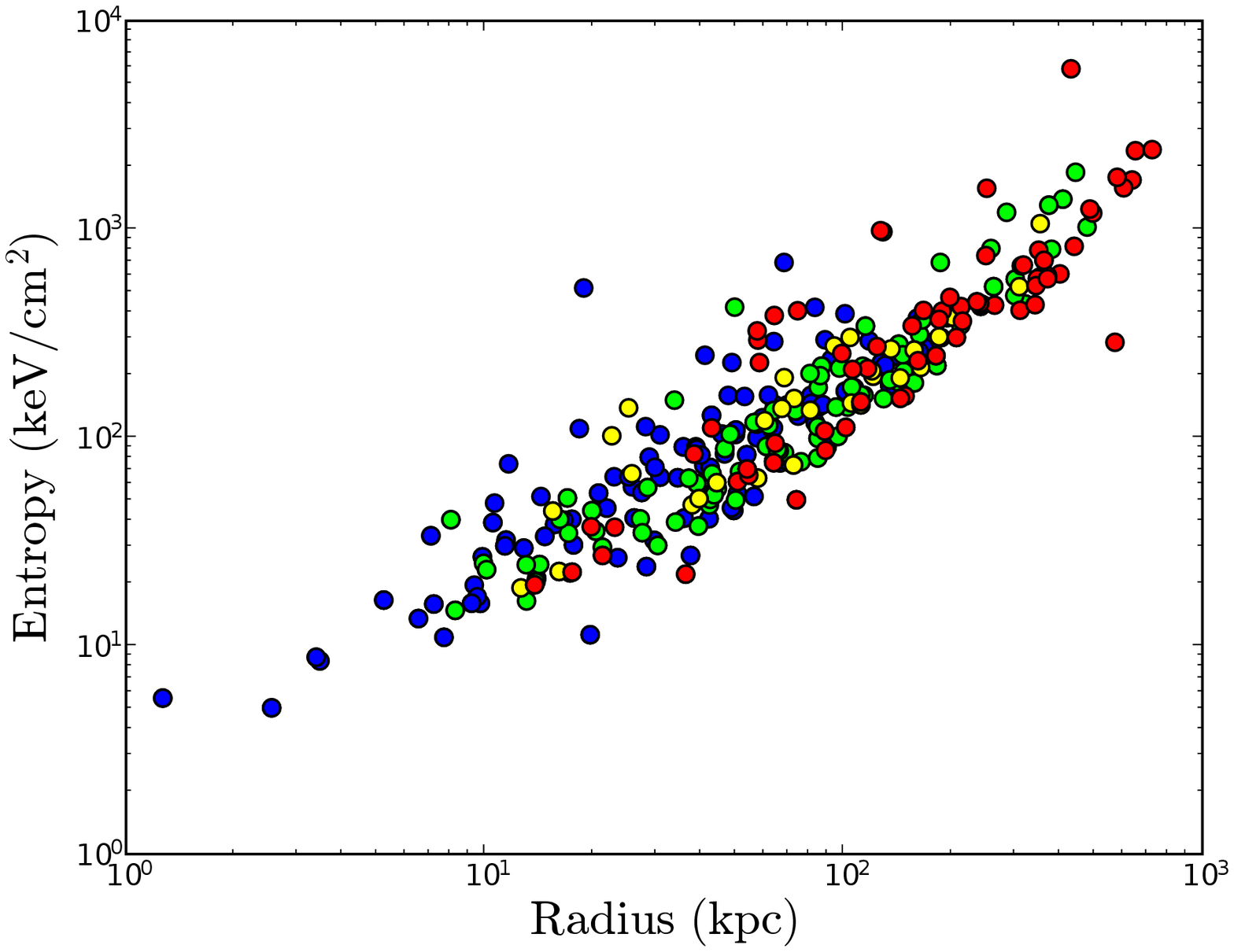}
\includegraphics[width=0.3\textwidth]{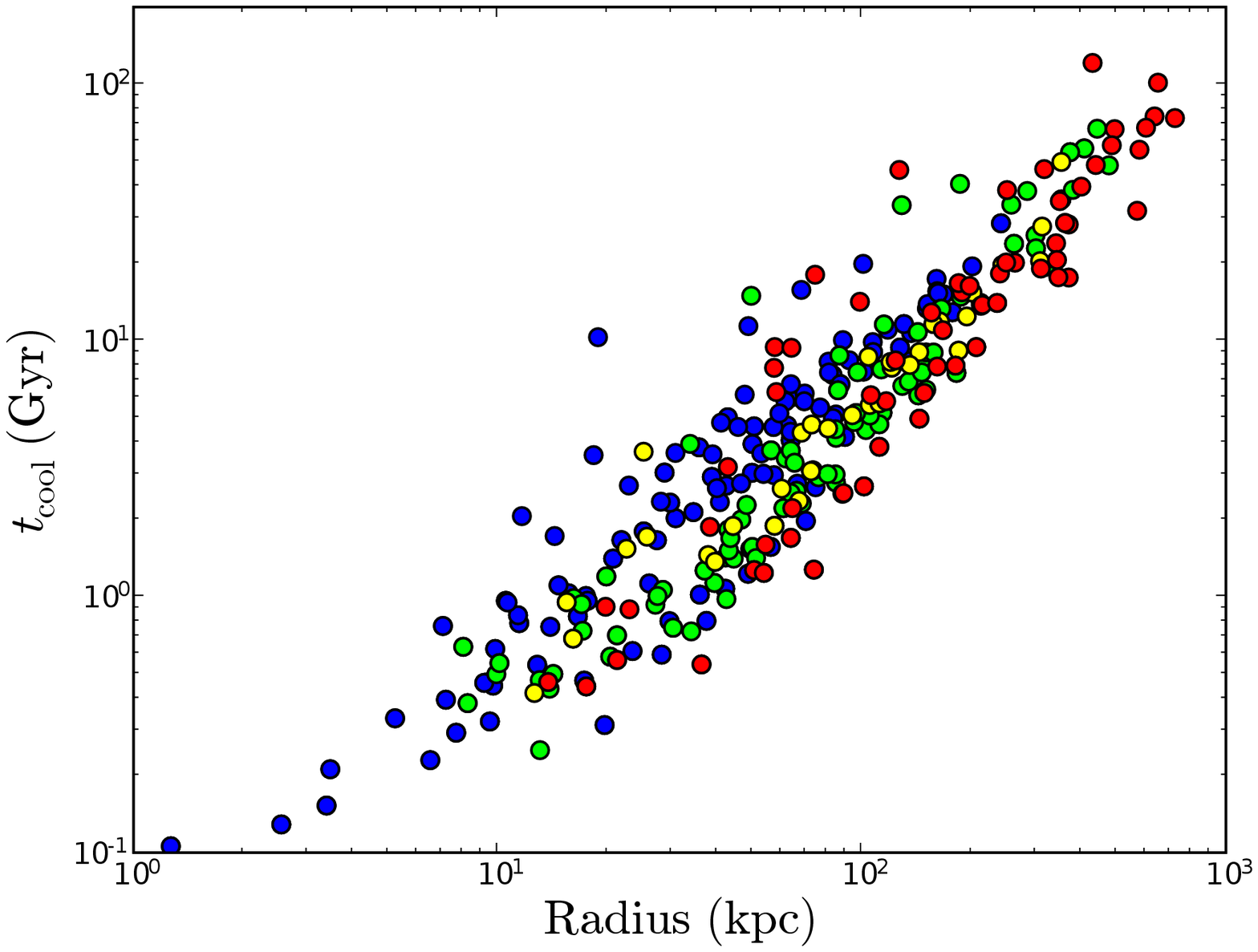}
\caption{\label{fig:combined_z} The distributions of the combined
  sample {\scshape left}: Temperature, {\scshape centre}: Entropy and
  {\scshape right}: Cooling time.  The colour scale shows the redshift
  of the clusters.  The redshifts have been binned into four bins with
  equal numbers of clusters (18) in each bin (blue $=z\leq 0.0521$,
  green $=0.0521<z\leq 0.161$,
  yellow $=0.161<z\leq 0.214$ and red $=z>0.214$).}
\end{figure*}

We have chosen to concentrate on the temperature, entropy and cooling
time profiles, and have not split the clusters on whether they
harbour clear bubbles or radio sources.  The profiles of all 71
clusters are shown in Fig. \ref{fig:combined_z} binned into four
redshift bins with equal numbers of clusters in each.

An initial glance at the temperature profiles gives the impression
that the lowest redshift clusters are cooler, with central
temperatures to below $1\kev$.  However, this is more likely to be a
result of resolution rather than redshift evolution.  The innermost
annulus is going to correspond to a smaller real radius in the nearby
clusters as their emission extends over a larger area of the sky and
they are also going to be brighter.  

The entropy profiles also vary between the lowest and highest
redshifts, however this is also most likely the result of resolution.  

The cooling time profiles, however, do show some change between the
redshift bins which is more likely to be the result of evolution.  For
a given radius in the cluster\footnote{We note that these profiles have
  {\it not} been scaled to make them truly cluster independent.  They do
  however align very well without any scaling applied} the lowest
redshift clusters have a longer cooling time than the highest redshift
clusters.  However, this may still be the result of resolution.  In
the high redshift clusters the inner regions of the cluster may all be
merged together in a single region.  This would reduce the average
cooling time of this bin.  Whereas in the nearby clusters, this region
would be split into many separate regions with those with shorter
cooling times occuring at smaller radii.  The bulk of the scatter in
this plot does reduce at higher radii, which is compatible with this argument.

\section{Conclusions}

We have extended our study of the effect of central AGN in clusters to
clusters with $z>0.1$, using the BCS.  There are only six clusters which
harbour clear bubbles, and on average they overcompensate for the
X-ray cooling.  These
clusters are combined with an updated analysis of those in the B55
cluster sample.  The average distance, as a fraction of the cooling radius,  out to which the bubbles can offset the X-ray
cooling from the combined sample is
$r_{\rm heat}/r_{\rm cool}=0.92\pm0.11$ for clusters where this
distance can be reliably determined.  Adding in all clusters with
bubbles results in $r_{\rm heat}/r_{\rm
  cool}=2.10\pm0.40$.

Although it appears as if the AGN bubbles can easily counteract the
X-ray cooling within the cooling radius, there is a selection effect,
especially at the higher redshifts, which means we are missing small
bubbles in the most distant clusters.  

Using the result from the BCS bubbles, we calculate the expected size
of bubbles within the clusters which harbour radio sources.  In three
cases (A1068, A1763 and A2204) the predicted bubble sizes and observed
radio emission are similar.  In some of the clusters the radio source
identified in the NVSS is offset to the cluster centre in VLA
observations, and so these may be background AGN rather than
associated with the cluster.  A category of clusters with
cool cores, radio sources and no clear bubbles exists in both the B55
and BCS samples. The lack of bubbles is explained by low
signal-to-noise ratio in the X-ray images of some but not all of these
objects.  

The distribution of the central cooling time of the clusters shows
that there are no clusters with bubbles which have $t_{\rm
  cool}>1.2\gyr$.  Also there are only 10/24 clusters with  $t_{\rm
  cool}>4.0\gyr$ which harbour a central radio source, only 5/47 clusters
with  $t_{\rm  cool}<4.0\gyr$ have no central radio source.

We investigated the evolution of cluster parameters with redshift and
over the range $0.0<z<0.4$ we do not find any significant change.  The
cooling time profiles show some variation but this can be explained
as being the result of the resolution differences as the distances
increase.  

\section*{Acknowledgements}

We thank Steve Allen, Roderick Johnstone, Jeremy Sanders, Paul
Alexander, Steve Rawlings and James Graham for
interesting discussions during the course of this work.  We thank the
referee for helpful comments in improving this manuscript.
The plots in this manuscript were created with {\it Veusz} and
{\scshape matplotlib}.

\bibliographystyle{mn2e} 
\bibliography{mn-jour,dunn2}

\begin{thebibliography}{}

\bibitem[\protect\citeauthoryear{{Adelman-McCarthy} \& {for the SDSS
  Collaboration}}{{Adelman-McCarthy} \& {for the SDSS
  Collaboration}}{2007}]{Adelman07}
{Adelman-McCarthy} J.~K.,  {for the SDSS Collaboration} 2007,
  arxiv.org/abs/0707.3413

\bibitem[\protect\citeauthoryear{{Arnaud}}{{Arnaud}}{1996}]{Arnaud96}
{Arnaud} K.~A.,  1996, in {Jacoby} G.~H.,  {Barnes} J.,  eds, ASP Conf. Ser.
  101: Astronomical Data Analysis Software and Systems V {XSPEC: The First Ten
  Years}.
p.~17

\bibitem[\protect\citeauthoryear{{Bauer}, {Fabian}, {Sanders}, {Allen} \&
  {Johnstone}}{{Bauer} et~al.}{2005}]{Bauer05}
{Bauer} F.~E.,  {Fabian} A.~C.,  {Sanders} J.~S.,  {Allen} S.~W.,
  {Johnstone} R.~M.,  2005, \mnras, 359, 1481

\bibitem[\protect\citeauthoryear{{B\^irzan}, {Rafferty}, {McNamara}, {Wise} \&
  {Nulsen}}{{B\^irzan} et~al.}{2004}]{Birzan04}
{B\^irzan} L.,  {Rafferty} D.~A.,  {McNamara} B.~R.,  {Wise} M.~W.,    {Nulsen}
  P.~E.~J.,  2004, \apj, 607, 800

\bibitem[\protect\citeauthoryear{{Blanton}, {Sarazin}, {McNamara} \&
  {Wise}}{{Blanton} et~al.}{2001}]{Blanton01}
{Blanton} E.~L.,  {Sarazin} C.~L.,  {McNamara} B.~R.,    {Wise} M.~W.,  2001,
  \apjl, 558, L15

\bibitem[\protect\citeauthoryear{{B{\"o}hringer}, {Voges}, {Fabian}, {Edge} \&
  {Neumann}}{{B{\"o}hringer} et~al.}{1993}]{Bohringer93}
{B{\"o}hringer} H.,  {Voges} W.,  {Fabian} A.~C.,  {Edge} A.~C.,    {Neumann}
  D.~M.,  1993, \mnras, 264, L25

\bibitem[\protect\citeauthoryear{{Burns}}{{Burns}}{1990}]{Burns90}
{Burns} J.~O.,  1990, \aj, 99, 14

\bibitem[\protect\citeauthoryear{{Churazov}, {Br{\" u}ggen}, {Kaiser}, {B{\"
  o}hringer} \& {Forman}}{{Churazov} et~al.}{2001}]{Churazov01}
{Churazov} E.,  {Br{\" u}ggen} M.,  {Kaiser} C.~R.,  {B{\" o}hringer} H.,
  {Forman} W.,  2001, \apj, 554, 261

\bibitem[\protect\citeauthoryear{{Churazov}, {Forman}, {Jones} \& {B{\"
  o}hringer}}{{Churazov} et~al.}{2000}]{Churazov00}
{Churazov} E.,  {Forman} W.,  {Jones} C.,    {B{\" o}hringer} H.,  2000, \aap,
  356, 788

\bibitem[\protect\citeauthoryear{{Churazov}, {Sunyaev}, {Forman} \&
  {B{\"o}hringer}}{{Churazov} et~al.}{2002}]{Churazov02}
{Churazov} E.,  {Sunyaev} R.,  {Forman} W.,    {B{\"o}hringer} H.,  2002,
  \mnras, 332, 729

\bibitem[\protect\citeauthoryear{{Donahue}, {Horner}, {Cavagnolo} \&
  {Voit}}{{Donahue} et~al.}{2006}]{Donahue06}
{Donahue} M.,  {Horner} D.~J.,  {Cavagnolo} K.~W.,    {Voit} G.~M.,  2006,
  \apj, 643, 730

\bibitem[\protect\citeauthoryear{{Dunn} \& {Fabian}}{{Dunn} \&
  {Fabian}}{2006}]{Dunn06f}
{Dunn} R.~J.~H.,  {Fabian} A.~C.,  2006, \mnras, 373, 959

\bibitem[\protect\citeauthoryear{{Dunn}, {Fabian} \& {Taylor}}{{Dunn}
  et~al.}{2005}]{Dunn05}
{Dunn} R.~J.~H.,  {Fabian} A.~C.,    {Taylor} G.~B.,  2005, \mnras, 364, 1343

\bibitem[\protect\citeauthoryear{{Ebeling}, {Edge}, {Allen}, {Crawford},
  {Fabian} \& {Huchra}}{{Ebeling} et~al.}{2000}]{Ebeling00}
{Ebeling} H.,  {Edge} A.~C.,  {Allen} S.~W.,  {Crawford} C.~S.,  {Fabian}
  A.~C.,    {Huchra} J.~P.,  2000, \mnras, 318, 333

\bibitem[\protect\citeauthoryear{{Ebeling}, {Edge}, {Bohringer}, {Allen},
  {Crawford}, {Fabian}, {Voges} \& {Huchra}}{{Ebeling}
  et~al.}{1998}]{Ebeling98}
{Ebeling} H.,  {Edge} A.~C.,  {Bohringer} H.,  {Allen} S.~W.,  {Crawford}
  C.~S.,  {Fabian} A.~C.,  {Voges} W.,    {Huchra} J.~P.,  1998, \mnras, 301,
  881

\bibitem[\protect\citeauthoryear{{Edge}, {Stewart}, {Fabian} \&
  {Arnaud}}{{Edge} et~al.}{1990}]{Edge90}
{Edge} A.~C.,  {Stewart} G.~C.,  {Fabian} A.~C.,    {Arnaud} K.~A.,  1990,
  \mnras, 245, 559

\bibitem[\protect\citeauthoryear{{Fabian}, {Sanders}, {Allen}, {Crawford},
  {Iwasawa}, {Johnstone}, {Schmidt} \& {Taylor}}{{Fabian}
  et~al.}{2003}]{Fabian03a}
{Fabian} A.~C.,  {Sanders} J.~S.,  {Allen} S.~W.,  {Crawford} C.~S.,  {Iwasawa}
  K.,  {Johnstone} R.~M.,  {Schmidt} R.~W.,    {Taylor} G.~B.,  2003, \mnras,
  344, L43

\bibitem[\protect\citeauthoryear{{Fabian}, {Sanders}, {Crawford}, {Conselice},
  {Gallagher} \& {Wyse}}{{Fabian} et~al.}{2003}]{Fabian03b}
{Fabian} A.~C.,  {Sanders} J.~S.,  {Crawford} C.~S.,  {Conselice} C.~J.,
  {Gallagher} J.~S.,    {Wyse} R.~F.~G.,  2003, \mnras, 344, L48

\bibitem[\protect\citeauthoryear{{Gull} \& {Northover}}{{Gull} \&
  {Northover}}{1973}]{Gull73}
{Gull} S.~F.,  {Northover} K.~J.~E.,  1973, \nat, 244, 80

\bibitem[\protect\citeauthoryear{{Johnstone}, {Allen}, {Fabian} \&
  {Sanders}}{{Johnstone} et~al.}{2002}]{Johnstone02}
{Johnstone} R.~M.,  {Allen} S.~W.,  {Fabian} A.~C.,    {Sanders} J.~S.,  2002,
  \mnras, 336, 299

\bibitem[\protect\citeauthoryear{{McNamara}, {Wise}, {Nulsen}, {David},
  {Sarazin}, {Bautz}, {Markevitch}, {Vikhlinin}, {Forman}, {Jones} \&
  {Harris}}{{McNamara} et~al.}{2000}]{McNamara00}
{McNamara} B.~R.,  {Wise} M.,  {Nulsen} P.~E.~J.,  {David} L.~P.,  {Sarazin}
  C.~L.,  {Bautz} M.,  {Markevitch} M.,  {Vikhlinin} A.,  {Forman} W.~R.,
  {Jones} C.,    {Harris} D.~E.,  2000, \apjl, 534, L135

\bibitem[\protect\citeauthoryear{{Mewe}, {Kaastra} \& {Leidahl}}{{Mewe}
  et~al.}{1995}]{Mewe95}
{Mewe} R.,  {Kaastra} J.~S.,    {Leidahl} D.~A.,  1995, Legacy, 6, 16

\bibitem[\protect\citeauthoryear{{Nulsen}, {Hambrick}, {McNamara}, {Rafferty},
  {Birzan}, {Wise} \& {David}}{{Nulsen} et~al.}{2005}]{Nulsen05}
{Nulsen} P.~E.~J.,  {Hambrick} D.~C.,  {McNamara} B.~R.,  {Rafferty} D.,
  {Birzan} L.,  {Wise} M.~W.,    {David} L.~P.,  2005, \apjl, 625, L9

\bibitem[\protect\citeauthoryear{{Nulsen}, {McNamara}, {Wise} \&
  {David}}{{Nulsen} et~al.}{2005}]{Nulsen04}
{Nulsen} P.~E.~J.,  {McNamara} B.~R.,  {Wise} M.~W.,    {David} L.~P.,  2005,
  \apj, 628, 629

\bibitem[\protect\citeauthoryear{{Peres}, {Fabian}, {Edge}, {Allen},
  {Johnstone} \& {White}}{{Peres} et~al.}{1998}]{Peres98}
{Peres} C.~B.,  {Fabian} A.~C.,  {Edge} A.~C.,  {Allen} S.~W.,  {Johnstone}
  R.~M.,    {White} D.~A.,  1998, \mnras, 298, 416

\bibitem[\protect\citeauthoryear{{Peterson} \& {Fabian}}{{Peterson} \&
  {Fabian}}{2006}]{Peterson06}
{Peterson} J.~R.,  {Fabian} A.~C.,  2006, Physics Reports, 427, 1

\bibitem[\protect\citeauthoryear{{Peterson}, {Kahn}, {Paerels}, {Kaastra},
  {Tamura}, {Bleeker}, {Ferrigno} \& {Jernigan}}{{Peterson}
  et~al.}{2003}]{Peterson03}
{Peterson} J.~R.,  {Kahn} S.~M.,  {Paerels} F.~B.~S.,  {Kaastra} J.~S.,
  {Tamura} T.,  {Bleeker} J.~A.~M.,  {Ferrigno} C.,    {Jernigan} J.~G.,  2003,
  \apj, 590, 207

\bibitem[\protect\citeauthoryear{{Rafferty}, {McNamara}, {Nulsen} \&
  {Wise}}{{Rafferty} et~al.}{2006}]{Rafferty06}
{Rafferty} D.~A.,  {McNamara} B.~R.,  {Nulsen} P.~E.~J.,    {Wise} M.~W.,
  2006, \apj, 652, 216

\bibitem[\protect\citeauthoryear{{Sanders} \& {Fabian}}{{Sanders} \&
  {Fabian}}{2002}]{Sanders02}
{Sanders} J.~S.,  {Fabian} A.~C.,  2002, \mnras, 331, 273

\bibitem[\protect\citeauthoryear{{Voigt} \& {Fabian}}{{Voigt} \&
  {Fabian}}{2004}]{Voigt04}
{Voigt} L.~M.,  {Fabian} A.~C.,  2004, \mnras, 347, 1130

\end{thebibliography}

\end{document}